\documentclass[journal]{IEEEtran}
\usepackage{cite}
\usepackage{amsmath,amssymb,amsfonts}
\usepackage{graphicx}
\usepackage{subcaption}
\usepackage{textcomp}
\usepackage{xcolor}
\usepackage{float}

\usepackage{multirow}   
\usepackage{makecell}   
\usepackage{booktabs}   
\usepackage[linesnumbered,ruled,vlined]{algorithm2e}
\usepackage{wrapfig}
\usepackage[colorlinks=true,linkcolor=blue,anchorcolor=blue, citecolor=blue, urlcolor=magenta]{hyperref}

\usepackage{utfsym}   
\usepackage{url} 
\usepackage{threeparttable}
\usepackage{enumitem}
\usepackage{tabularx}
\usepackage{array}
\usepackage{courier}
\usepackage{fancyhdr}

\fancypagestyle{firstpage}{

  \fancyhead[C]{This paper is published in IEEE Transactions on Computer-Aided Design of Integrated Circuits and Systems (\url{https://ieeexplore.ieee.org/document/10750525})} 
  \fancyfoot[C]{}
}

\begin{document}

\title{PIMCOMP: An End-to-End DNN Compiler for Processing-In-Memory Accelerators}

\author{Xiaotian Sun, Xinyu Wang, Wanqian Li, 
        Yinhe~Han,~\IEEEmembership{Member,~IEEE,}
        and~Xiaoming~Chen,~\IEEEmembership{Member,~IEEE}
        \thanks{This work was supported in part by Strategic Priority Research Program of CAS under Grant XDB44000000, in part by National Natural Science Foundation of China under Grant 62122076, Grant 62025404, and Grant 62488101, in part by Key Research Program of Frontier Sciences, CAS under Grant ZDBS-LY-JSC012, and in part by Youth Innovation Promotion Association CAS. \textit{(Corresponding author: X. Chen)}}
\thanks{X. Sun, X. Wang,  W. Li, Y. Han, and X. Chen are with the Institute of Computing Technology, Chinese Academy of Sciences, Beijing 100190, China, and also with the University of Chinese Academy of Sciences, Beijing 100190, China (e-mail: sunxiaotian21s@ict.ac.cn; wangxinyu22s@ict.ac.cn; liwanqian20s@ict.ac.cn; yinhes@ict.ac.cn; chenxiaoming@ict.ac.cn).}
}


\maketitle

\thispagestyle{firstpage}

\begin{abstract}

In the past decade, various processing-in-memory (PIM) accelerators based on various devices, micro-architectures, and interfaces have been proposed to accelerate deep neural networks (DNNs). How to deploy DNNs onto PIM-based accelerators is the key to explore  PIM's high performance and energy efficiency. The scale of DNN models, the diversity of PIM accelerators, and the complexity of deployment are far beyond the human deployment capability. Hence, an automatic deployment methodology is indispensable. In this work, we propose PIMCOMP, an end-to-end DNN compiler tailored for PIM accelerators, achieving efficient deployment of DNN models on PIM hardware. PIMCOMP can adapt to various PIM architectures by using an abstract configurable PIM accelerator template with a set of pseudo-instructions, which is a high-level abstraction of the hardware's fundamental functionalities. Through a generic multi-level optimization framework, PIMCOMP realizes an end-to-end conversion from a high-level DNN description to pseudo-instructions, which can be further converted to specific hardware intrinsics/primitives. The compilation addresses two critical issues in PIM-accelerated inference from a system perspective: resource utilization and dataflow scheduling. PIMCOMP adopts a flexible unfolding format to reshape and partition  convolutional layers, adopts a weight-layout guided computation-storage-mapping approach to enhance resource utilization, and balances the system's computation, memory access, and communication characteristics. For dataflow scheduling, we design two scheduling algorithms with different inter-layer pipeline granularities to support varying application scenarios while ensuring high computational parallelism. Experiments demonstrate that PIMCOMP improves throughput, latency, and energy efficiency across various architectures.
PIMCOMP is open-sourced at \url{https://github.com/sunxt99/PIMCOMP-NN}.
\end{abstract}


\section{Introduction}

\IEEEPARstart{D}{EEP} neural networks (DNNs), with their powerful feature extraction and classification abilities, can excellently perform various intelligent tasks.
According to neural scaling laws, the size and data of neural network models are continuously increasing to achieve better performance. Researchers have proposed a series of special-purpose accelerators (e.g.,~\cite{EIE,DaDianNao}) to accommodate the models and speed up the inference process to cope with the ever-expanding neural networks. However, these accelerators are facing the memory wall challenge~\cite{MemoryWall} as they suffer from high-cost data movement between memory and processing elements and encounter obstacles in enhancing energy efficiency.

Processing-in-memory (PIM) is expected to overcome the memory wall challenge as it binds data and computation, thereby circumventing the need for data movement. Among various PIM implementations, the resistive random-access memory (RRAM) has high density, low latency, and is easy to integrate with the CMOS technology~\cite{RRAM}, offering a broad application prospect. RRAM cells are usually integrated into a crossbar array, which can perform a matrix-vector multiplication (MVM) in $O(1)$ time. The weights are programmed to be the conductances of the cells. The input activations are converted to voltages by digital-to-analog converters (DACs). 
According to the Kirchhoff's law, the output currents reflect the product of the weights (a matrix) and the inputs (a vector). Since the crossbar array operates in the analog domain, peripheral circuits such as DACs and analog-to-digital converters (ADCs) 
are needed to convert the signals. Due to the high memory density, parallel in-situ computing properties,  elimination of weight movement, and the crossbar array formed by PIM devices can potentially meet the storage and computation requirements of DNNs. Crossbar arrays can also be constructed with other devices, such as ferroelectric field-effect transistors, magnetic random-access memories, phase-change memories, and even volatile static random-access memories. Based on this principle, previous works (e.g.,~\cite{ISAAC, PRIME, Pipelayer, FeFET, MRAM, PCM}) have designed a series of  crossbar array-based DNN accelerators.

These accelerators contain thousands of or more crossbar arrays, which challenge the deployment of DNN models due to the vast hardware scale. In addition, DNN models with different sizes and topologies demand meticulous attention for resource utilization, data scheduling, and memory optimization during deployment. Due to the different micro-architectures of various accelerators, it is uneconomical and unrealistic to manually design the deployment schemes for each DNN model on each accelerator as in previous works~\cite{ISAAC, PRIME, Pipelayer, FeFET, MRAM, PCM}. Therefore, a compiler that can adapt to various PIM architectures and automatically complete DNN model deployment is indispensable to improve the usability of PIM accelerators, which also helps build a PIM ecosystem~\cite{Challenge}.

To bridge DNN models and PIM accelerators, the compiler needs to be designed by considering the following aspects to make it universal, flexible, and efficient.
\begin{itemize}
    \item \textbf{For hardware:} various PIM-based DNN accelerators have emerged, and it is cumbersome to design a specific compiler for each accelerator. Therefore, the compiler needs to be built on a high-level hardware abstraction, which can be lowered to specific PIM architectures. In addition, the compiler needs to be flexible and configurable to facilitate users quickly deploying DNNs on PIM accelerators with different scales of hardware resources.
    \item \textbf{For software:} the compiler should support a variety of DNN workloads. Besides, it should apply to different application scenarios, such as meeting low-latency or high-throughput requirements. Above all, the compiler should automatically complete the deployment, that is, transparently perform model reading, weight mapping, and output collecting without user intervention.
    \item \textbf{System-level optimization:} the compiler needs to handle resource allocation and dataflow scheduling effectively to unleash the hardware potential. For resource allocation, the compiler should make full use of the PIM resources to boost performance while balancing computing, memory access, and communication. For dataflow scheduling, the compiler should rapidly generate instruction streams for different scenarios and optimize the system performance bottlenecks. In addition, the compiler ought to have a profiler that provides an effective evaluation to steer the iterative performance optimization.
\end{itemize}

To address the challenges faced by PIM accelerators during DNN deployment, this paper proposes \emph{PIMCOMP}, an end-to-end DNN compiler for PIM accelerators. A previous version of this work was published in~\cite{PIMCOMP}, which provides an optimization scheme for resource allocation, task mapping, and pipeline dataflow through four stages: layer partitioning, weight replication, core mapping, and dataflow scheduling. The previous work offers a methodology that focuses on some deployment-specific issues, but falls short of meeting the compiler requirements stated above, as it lacks support for end-to-end compilation, integration of different compilation stages, and system-level optimizations. Nevertheless, the previous work lays out a design blueprint and algorithmic guidance for this work.

In this paper, we first review and contrast the existing PIM compilers. Then, we present the overall architecture and implementation details of \emph{PIMCOMP}. \emph{PIMCOMP} is a practical compiler that builds on Ref.~\cite{PIMCOMP} and enhances hardware compatibility, software support, and holistic performance. The new contributions of this paper are summarized as follows.
\begin{enumerate}
    \item We propose an end-to-end DNN compiler tailored for PIM accelerators, encompassing the frontend, optimizer, and backend, capable of inferring an entire DNN model. We design two pipelines to accommodate diverse application scenarios. We build a profiler to accomplish a comprehensive performance assessment for iterative compilation space optimizations.
    \item We present a PIM accelerator abstraction adaptable to various academic accelerator designs. The  accelerator abstraction comprises a hardware template with rich configurability, a set of pseudo-instructions fully abstracting the hardware's fundamental functionalities, and user-specified hardware execution patterns.
    \item We introduce three-stage optimizations to complete the deployment of DNNs on PIM accelerators. In the layer partitioning stage, we propose array groups as the basic programming unit and utilize a flexible unfolding format to meet various resource demands. In the layout-computation mapping stage, we propose a weight-layout guided computation-storage-mapping method, utilizing genetic algorithms to optimize weight replication and layout and adaptively allocating computational tasks. In the dataflow scheduling stage, we propose scheduling algorithms maintaining high parallelism tailored to two pipelines.
    \item We evaluate \emph{PIMCOMP}'s end-to-end deployment capability on three different architectures. The experimental results show that \emph{PIMCOMP} achieves promising improvements in throughput, latency, and energy consumption compared with previous works.
\end{enumerate}

\section{Compilation Tools for PIM}
\label{Sec::Tools}

\textcolor{black}{Previous works have explored the end-to-end conversion from high-level programs to low-level code \cite{TVM,ETE1,ETE2}, incorporating optimizations during the progressive lowering. However, these approaches designed for CPUs and GPUs with separate storage and computation are not well-suited to PIM, due to its characteristic of integrating storage and computation. Furthermore, the high parallelism enabled by the PIM crossbar array reduces the effectiveness of conventional optimization strategies, which are tailored for multi-level loops.}

Consequently, specialized compilation tools for PIM have been developed \cite{POLY,CODE,SONGC,TCCIM,TDOCIM,OCC,PUMA,CINM}.
\textcolor{black}{Among them, \emph{Polyhedral}~\cite{POLY}, \emph{Co-Design}~\cite{CODE}, \emph{SongC}~\cite{SONGC}, and \emph{PIMCOMP} focus on DNNs, while \emph{TC-CIM}~\cite{TCCIM}, \emph{TDO-CIM}~\cite{TDOCIM}, \emph{OCC}~\cite{OCC}, and \emph{PUMA}~\cite{PUMA} target a broader range of machine learning algorithms. In addition to these, \emph{CINM}~\cite{CINM} supports more general applications, such as time series analysis.}
\textcolor{black}{With DNNs as the target application, Table~\ref{Tab::Review} summarizes these compilation tools from the following perspectives.}

\begin{table*}[t]
\centering
\caption{Summary and comparison of different compiling tools for PIM, \textcolor{black}{focusing on DNN applications}.}
\label{Tab::Review}
\setlength{\tabcolsep}{4pt}
\begin{threeparttable}
\begin{tabular}{|c|c|c|c|c|c|c|}
\hline
           & Design philosophy  &   Sche. gran.    & \textcolor{black}{E2E for DNN}  &  HW conﬁg.        & App. diver.         & Sys. opt.  \\ \hline

TC-CIM\cite{TCCIM}  & Naive computation-storage-binding & \textcolor{black}{Layer-wise MVM oper.} & \usym{2715} & Low & Low & -  \\ \hline
TDO-CIM\cite{TDOCIM} & Naive computation-storage-binding & \textcolor{black}{Layer-wise MVM oper.} & \usym{2715} &  Low & Low & -  \\ \hline
CINM\cite{CINM}  & Naive computation-storage-binding & \textcolor{black}{Layer-wise MVM oper.}  & \usym{2715} &  \textcolor{black}{High} & Low  & P\tnote{1}  \\ \hline
OCC\cite{OCC}    & Naive computation-storage-binding & \textcolor{black}{Layer-wise MVM oper.} & \usym{2715} &  Low & Low  & P  \\ \hline
Polyhedral\cite{POLY} &  Naive computation-storage-binding & \textcolor{black}{Layer-wise MVM oper.} & \usym{2715} &  Medium & Medium & P/\textcolor{black}{H}\tnote{2}  \\ \hline
Co-Design\cite{CODE}  & Naive computation-storage-binding & \textcolor{black}{CG-based MVM oper.} & \usym{2713} & Medium & Medium & P/\textcolor{black}{H}  \\ \hline
PUMA\cite{PUMA}  & Naive computation-storage-binding & \textcolor{black}{CG-based MVM oper.} & \usym{2715}  & Medium & Medium & P/M\tnote{3}  \\ \hline
SongC\cite{SONGC}  & Storage-computation-disentangled & \textcolor{black}{Layer-wise MVM oper.}  & \usym{2715} & Medium & Low & P/M  \\ \hline
\textbf{PIMCOMP} & Weight-layout guided computation-storage-mapping & \textcolor{black}{CG-based Conv. oper.} & \usym{2713} & High & High & P/\textcolor{black}{H}/M  \\ \hline

\end{tabular}
    \begin{tablenotes}
    \item[] $^1$P: performance. \quad $^2$\textcolor{black}{H: hardware resource utilization.}  \quad $^3$M: memory orchestration
    \end{tablenotes}
\end{threeparttable}
\end{table*}

(1) \textbf{Design philosophy}: The primary concept of PIM is to bind computation and data, namely, performing computations where data reside. Thus, the data layout is crucial. For accelerating DNNs on PIM, there is a crucial strategy called \textit{weight replication}~\cite{ISAAC, Pipelayer, Replication}. By replicating a layer's weights by $R$ times, the computational parallelism of that layer is increased by $R$ times so the performance can be boosted by $R$ times at most. \emph{TC-CIM}~\cite{TCCIM}, \emph{TDO-CIM}~\cite{TDOCIM}, \emph{CINM}~\cite{CINM}, \emph{OCC}~\cite{OCC}, \emph{Polyhedral}~\cite{POLY}, \emph{Co-Design}~\cite{CODE}, and \emph{PUMA}~\cite{PUMA} do not consider the interaction between weight replication and weight layout, and simply map computational tasks to PIM arrays. This approach may lead to suboptimal performance and resource under-utilization. \emph{SongC}~\cite{SONGC} considers a storage-computation-disentangled approach, which resembles non-PIM accelerators, resulting in additional overhead due to weight movement. \emph{PIMCOMP} employs a weight-layout guided computation-storage-mapping method, which carefully determines the weight layout for all layers under weight replication consideration and subsequently adaptively maps computational tasks to PIM arrays based on the weight replication and layout. This approach not only fully utilizes hardware resources, but more importantly, eliminates mismatch between the requirements of computation and storage.

(2) \textbf{Schedule granularity}: \textcolor{black}{\emph{TC-CIM}~\cite{TCCIM}, \emph{TDO-CIM}~\cite{TDOCIM}, \emph{CINM}~\cite{CINM}, \emph{OCC}~\cite{OCC}, \emph{Polyhedral}~\cite{POLY}, and \emph{SongC}~\cite{SONGC} convert convolution layers into fundamental MVM operators in a layer-by-layer manner.}
This is a natural idea, but, such a coarse-grained granularity presents a challenge: when multiple layers are mapped to a core, they need to be computed sequentially, leading to resource under-utilization. On the contrary, in \emph{Co-Design}~\cite{CODE} and \emph{PUMA}~\cite{PUMA}, scheduling is performed at the granularity of MVM operators across the model, incurring significant overhead in parallelism analysis and operation linearization, resulting in prolonged compilation time. To fully exploit hardware resources and facilitate efficient scheduling, \emph{PIMCOMP} adopts a convolution operator\footnote{A convolution operator is the computation associated with a sliding window on the input feature maps and all filters, producing an output element on each output feature map.}-level scheduling strategy, ensuring that operators without structure and data conflicts can be executed in parallel, thereby enhancing operator-level parallelism.

(3) \textbf{End-to-end for DNN}: An end-to-end DNN compiler should act as a black box that compiles a user-given DNN model into logically correct instructions or operations. Two key features should be realized for an end-to-end compiler. First, the compiler should not ask users to rewrite the model. \textcolor{black}{\emph{OCC}~\cite{OCC}, \emph{CINM}~\cite{CINM},} \emph{Co-Design}~\cite{CODE}, \emph{SongC}~\cite{SONGC}, and \emph{PIMCOMP} eliminate the need for rewriting the models, while the others require users to manually convert the model into the special input format that the compiler can accept. 
Second, the compiler should have the capability to compile an entire DNN model. \emph{TC-CIM}~\cite{TCCIM}, \emph{TDO-CIM}~\cite{TDOCIM}, \emph{CINM}~\cite{CINM}, \emph{OCC}~\cite{OCC}, and \emph{SongC}~\cite{SONGC} can only compile single layers or blocks within the networks.
\emph{Co-Design}~\cite{CODE} and \emph{PIMCOMP} both use the ONNX format\footnote{\url{https://onnx.ai}.} as input to avoid rewriting and provide mechanisms to support compilation of the entire DNN model.

(4) \textbf{Hardware configurability:} \emph{TC-CIM}~\cite{TCCIM}, \emph{TDO-CIM}~\cite{TDOCIM}, and \emph{OCC}~\cite{OCC} assume relatively fixed hardware structures, which limits the hardware configurability. \textcolor{black}{\emph{CINM}~\cite{CINM} provides hardware abstraction, enabling support for various hardware platforms.} The other compilers support flexible hardware structures. Among them, \emph{PIMCOMP} supports a rich set of configurable parameters of various hardware components, such as the scale of cores, the size of PIM crossbar arrays, and bandwidths of different levels of memories and network-on-chip (NoC).
    
(5) \textbf{Application-scenario diversity:} Real-world tasks have different performance requirements. For example, cloud devices need high throughput, while mobile devices need low latency. Since PIM accelerators store all weights in crossbar arrays, compilers can adjust the inter-layer pipeline granularities to suit various application scenarios. \emph{TC-CIM}~\cite{TCCIM}, \emph{TDO-CIM}~\cite{TDOCIM}, \emph{CINM}~\cite{CINM}, \emph{OCC}~\cite{OCC}, and \emph{SongC}~\cite{SONGC} can only perform layer-by-layer deployment for a single sample. \emph{Polyhedral}~\cite{POLY} and \emph{CO-Design}~\cite{CODE} implement sample-granularity pipelines to adapt to high-throughput scenarios. \emph{PUMA}~\cite{PUMA} designs an inter-core pipeline at the MVM granularity to reduce latency. \emph{PIMCOMP} designs specifically optimized scheduling algorithms for two different scenarios, enhancing the applicability of PIM hardware.

(6) \textbf{System optimization:} To fully exploit the capability of PIM hardware, the compiler should explore various optimization opportunities, such as performance, hardware resource utilization, and memory orchestration. \emph{TC-CIM}~\cite{TCCIM} and \emph{TDO-CIM}~\cite{TDOCIM} identify and offload suitable operations to PIM hardware but lack optimization measures. \emph{CINM}~\cite{CINM} and \emph{OCC}~\cite{OCC} apply loop unrolling on this basis to enhance the parallelism. \emph{Polyhedral}~\cite{POLY} adopts a polyhedral model to optimize inference performance and increases the resource utilization by weight replication. \emph{Co-Design}~\cite{CODE} and \emph{PUMA}~\cite{PUMA} employ graph partitioning algorithms to minimize data transmission between cores. For memory orchestration, \emph{PUMA}~\cite{PUMA} reuses memory space and \emph{SongC}~\cite{SONGC} compresses redundant memory blocks. \emph{PIMCOMP} leverages resources and boosts performance through automated layout-computation mapping (Section~\ref{Sec::Mapping}) and devises runtime memory management strategies for low-latency pipelining (Section~\ref{Sec::Dataflow}).

\section{PIM Accelerator Abstraction}
\label{Sec::Architecture}

To make \emph{PIMCOMP} compatible with various PIM accelerators, we propose a high-level abstraction that includes an architecture template, pseudo-instructions as a software-hardware interface, and configurable execution patterns. This abstraction acts as a hardware abstraction layer~\cite{HAL}, isolating hardware differences from the compiler to improve \emph{PIMCOMP}'s compatibility and portability. In this section, we first introduce the architecture template, which macroscopically defines the functionality and organization of the fundamental modules. 
Then, we present the software-hardware interface serving as the interaction layer between  DNN models and the abstract architecture.
Based on this, we propose configurable execution patterns to specify the execution logic of the abstract hardware, facilitating the compilation adapting to the actual hardware execution patterns.

\begin{figure}[t]
  \centering
  \includegraphics[width=1.0\linewidth]{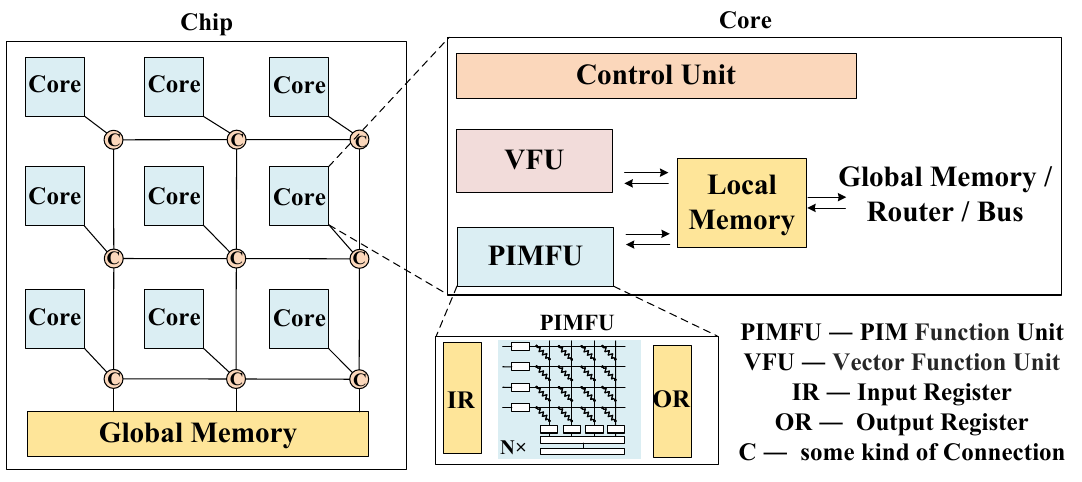}\\
  \caption{Architecture template.}
  \label{Fig::Architecture}
\end{figure}

\subsection{Architecture Template}
The architecture template is a representative multi-level configurable and scalable architecture derived from previous PIM accelerators~\cite{ISAAC,PUMA,AtomLayer,PUMA}. Fig.~\ref{Fig::Architecture} illustrates the architecture template comprising chips, cores, PIM function units (PIMFUs), crossbar arrays, and multi-level storage.
\textcolor{black}{Given our commitment to expanding the applicability of \emph{PIMCOMP}, this template does not introduce additional innovations in hardware design. Its distinguishing feature lies in the extensive configurability at various levels.
The configurable parameters, listed in Table~\ref{Tab::Parameter}, facilitate the instantiation of the template into a tailored accelerator.} The \textbf{accelerator} architecture consists of multiple chips interconnected by off-chip connections, such as PCIe.

Each \textbf{chip} contains multiple cores connected to the global memory. These cores are interconnected by configurable methods, such as buses, NoCs, or shared memories. The cores perform parallel computation and can communicate by synchronous or asynchronous mechanisms.

The \textbf{core}  
is the computing unit that performs MVMs with a PIMFU and vector computations with vector function units (VFUs), with a  local scratchpad memory to store intermediate results. Users can customize the VFU to perform different vector operations, such as result accumulation with an adder tree~\cite{MNSIM2}, max pooling with max pool units~\cite{ISAAC}, activation with activation units~\cite{ISAAC}, or various vector operations with integrated multi-function vector units~\cite{PUMA}. The local memory represents an abstraction of all storage units within a core, facilitating data exchange with both PIMFUs and VFUs while also storing data transmitted between cores. Moreover, the local memory can handle padding, concatenation, and splitting operations. The presence of the local memory enables the compiler to schedule the dataflow, thereby achieving an on-chip pipeline flexibly.

\begin{table}[t]
\caption{Configurable parameters for the architecture template.}
\label{Tab::Parameter}
\centering
\setlength{\tabcolsep}{1pt}
\begin{tabular}{|c|c||c|c|}
\hline
Level                        & Parameter       & Level                                                                     & Parameter              \\ \hline
\multirow{3}{*}{Accelerator} & \#Chip in X       & \multirow{5}{*}{Core}                                                     & VFU functionality        \\ \cline{2-2} \cline{4-4} 
                             & \#Chip in Y       &                                                                           & \#VFU           \\ \cline{2-2} \cline{4-4} 
                             & Off-chip bandwidth     &                                                                           & LocalMem bandwidth            \\ \cline{1-2} \cline{4-4} 
\multirow{7}{*}{Chip}        & \#Core in X      &                                                                           & LocalMem size          \\ \cline{2-2} \cline{4-4} 
                             & \#Core in Y       &                                                                           & Instruction execution model  \\ \cline{2-4} 
                             & Connection type & \multirow{3}{*}{PIMFU}                                                    & \#Crossbar in X          \\ \cline{2-2} \cline{4-4} 
                             & Connection bandwidth   &                                                                           & \#Crossbar in Y          \\ \cline{2-2} \cline{4-4} 
                             & Comm. mechanism &                                                                           & Management granularity \\ \cline{2-4} 
                             & GlobalMem bandwidth    & \multirow{2}{*}{\begin{tabular}[c]{@{}c@{}}Crossbar\\ Array\end{tabular}} & Crossbar array size    \\ \cline{2-2} \cline{4-4} 
                             & GlobalMem size  &                                                                           & Cell precision         \\ \hline
\end{tabular}
\end{table}

\begin{table*}[t]
\caption{Software-hardware interface of pseudo-instructions.}
\label{Tab::Interface}
\centering
\setlength{\tabcolsep}{2.8pt}
\begin{tabular}{|m{1.8cm}|m{3cm}|m{3cm}|m{8.8cm}|}
\hline
Operation type & Pseudo-instruction & Target & Description \\ \hline
\multirow{2}{*}{Computation} & mvm(\textcolor{teal}{\%idx}, \textcolor{teal}{\%dst}, \textcolor{teal}{\%src}, \textcolor{teal}{\%len}) & PIMFU & Array group with index \textcolor{teal}{\%idx} reads data of length \textcolor{teal}{\%len} at \textcolor{teal}{\%src} in LocalMem, performs MVM, and then writes the result back to LocalMem at \textcolor{teal}{\%dst}. \\ \cline{2-4} 
 & vec(\textcolor{teal}{\%op}, \textcolor{teal}{\%dst}, \textcolor{teal}{\%src1}, \textcolor{teal}{\%src2}, \textcolor{teal}{\%len}) & VFU & VFU performs vector calculations on data of length \textcolor{teal}{\%len} at \textcolor{teal}{\%src1} and \textcolor{teal}{\%src2} in LocalMem, and writes the result to \textcolor{teal}{\%dst}. \textcolor{teal}{\%op} identifies the operations. \\ \hline
\multirow{4}{*}{Memory} & copy(\textcolor{teal}{\%dst}, \textcolor{teal}{\%src}, \textcolor{teal}{\%len}) & LocalMem & Copy the data  at \textcolor{teal}{\%src} of length \textcolor{teal}{\%len} to \textcolor{teal}{\%dst} in LocalMem \\ \cline{2-4} 
 & write(\textcolor{teal}{\%dst}, \textcolor{teal}{\%len}, \textcolor{teal}{\%imm}) & LocalMem & Write \textcolor{teal}{\%imm} to data of length \textcolor{teal}{\%len} at \textcolor{teal}{\%dst} in LocalMem  \\ \cline{2-4} 
 & load(\textcolor{teal}{\%dst}, \textcolor{teal}{\%src}, \textcolor{teal}{\%len}) & GlobalMem\textrightarrow LocalMem & Copy data of length \textcolor{teal}{\%len} from \textcolor{teal}{\%src} in GloalMem to \textcolor{teal}{\%dst} in LocalMem \\ \cline{2-4} 
 & store(\textcolor{teal}{\%dst}, \textcolor{teal}{\%src}, \textcolor{teal}{\%len}) & LocalMem\textrightarrow GlobalMem & Write data of length \textcolor{teal}{\%len} from \textcolor{teal}{\%src} in LocalMem to \textcolor{teal}{\%dst} in GlobalMem \\ \hline
\multirow{2}{*}{Communication} & send(\textcolor{teal}{\%idx}, \textcolor{teal}{\%src}, \textcolor{teal}{\%len}) & Core $\rightarrow$ Core & Transmit data of length \textcolor{teal}{\%len} from \textcolor{teal}{\%src} in LocalMem to Core \textcolor{teal}{\%idx} \\ \cline{2-4} 
 & recv(\textcolor{teal}{\%idx}, \textcolor{teal}{\%dst}, \textcolor{teal}{\%len}) & Core $\rightarrow$ Core & Receive data from Core \textcolor{teal}{\%idx} and write to LocalMem at \textcolor{teal}{\%dst} of length \textcolor{teal}{\%len} \\ \hline
\end{tabular}
\end{table*}

The \textbf{PIMFU} is a collection of PIM crossbar arrays and their peripheral circuits intended to process MVMs. Due to the limited precision of crossbar array cells, high-precision weights usually need multiple crossbar arrays to store collaboratively~\cite{MNSIM2}. Similarly, a signed weight may need two crossbar arrays to store the positive and negative parts separately~\cite{PRIME}. To support multi-precision deployment~\cite{MixedPresion} and simplify the compiler design, we use a logical array with each cell storing a signed weight with arbitrary precision to replace multiple physical arrays that form the same weight jointly in the optimizer stage. The compiler records the mapping relationships between the logical and physical arrays and performs weight-bit splitting in the backend stage based on this information. Moreover, the arrays are loosely coupled in PIMFU, meaning that their organization is not specified. To adapt to different control granularities of various PIM accelerators, we design a management granularity to represent the basic control unit of PIMFU, which ranges from fine to coarse, including a single array, a specific number of arrays, and all arrays in PIMFU.

Users can configure the size and cell precision of the \textbf{crossbar array} in PIMFU. By configuring different array sizes for different cores, a mixed-size deployment can be realized~\cite{MixedSize}. 

\subsection{Software-Hardware Interface of Pseudo-Instructions}

\textcolor{black}{The software-hardware interface is exposed at the core level of the hardware template because 
the core is responsible for managing and offloading computation, memory access, and data transmission tasks.}
By programming each core, the compiler can drive the entire accelerator for DNN inference.

We design a set of pseudo-instructions as the software-hardware interface as listed in Table~\ref{Tab::Interface}, where each pseudo-instruction abstracts a fundamental functionality of the core. For example, \texttt{mvm} controls the array group (the concept of array group will be introduced in Section~\ref{Sec::Partitioning}) identified by \texttt{\%idx} to perform an MVM operation. Within \texttt{vec}, the \texttt{\%op} represents the identification of various vector operations, such as activation operations of a single vector, arithmetic operations between two vectors, and comparative operations, among others.

Previous studies have suggested different application program interfaces (APIs) or instruction set architectures (ISAs) for PIM accelerators, but they are significantly different from the  interface presented in this paper. \emph{TDO-CIM}~\cite{TDOCIM}  develops a PIM runtime library to invoke the PIM co-processor. However, its management only covers data transmission between the host and device and the launch of the general matrix multiplication (GEMM) kernel. This coarse-grained management fails to exploit the full potential of PIM hardware and is inadequate for achieving sophisticated scheduling, such as fine-grained inter-layer pipelines and data reuse across convolution operators. \emph{CINM}~\cite{CINM}, \emph{OCC}~\cite{OCC}, and \emph{Polyhedral}~\cite{POLY} face the same issue.  \emph{Co-Design}~\cite{CODE} and \emph{PUMA}~\cite{PUMA} design their ISAs for the programmable hardware they propose. However, their ISAs are only compatible with specific architectures, lacking the generality and universality required for broader applicability.

\subsection{Execution Patterns}

Some configurable parameters of the hardware template pertain to the execution characteristics of the accelerator. To ensure the compilation results match the underlying hardware's operational logic, \emph{PIMCOMP} appropriately adjusts and optimizes the pseudo-instruction order based on the user-specified hardware execution patterns. We illustrate \emph{PIMCOMP}'s efforts using the instruction execution model and communication mechanism in Table~\ref{Tab::Parameter} as examples.

\begin{figure}[t]
  \centering
  \includegraphics[width=0.7\linewidth]{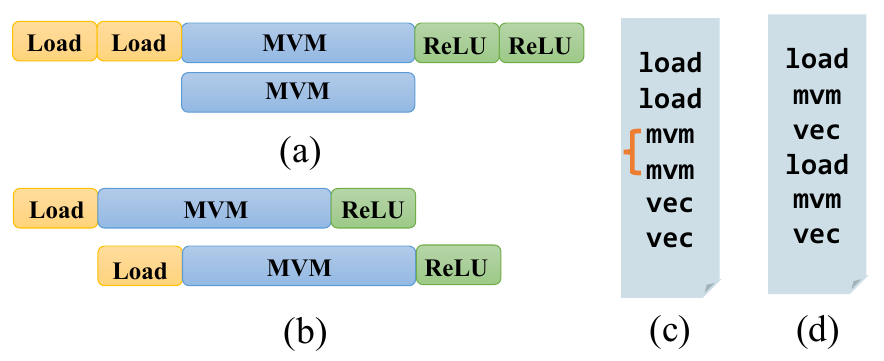}\\
  \caption{Example of adjustment based on instruction execution model. (a) and (c) depict the computational schematic and compilation results for in-order instruction execution model, respectively. (b) and (d) represent the computational schematic and compilation results for out-of-oder instruction execution model, respectively.}
  \label{Fig::Model-1}
\end{figure}

\textbf{(1) Instruction execution model}: For accelerators targeted by \emph{Co-Design}~\cite{CODE} and \emph{PUMA}~\cite{PUMA}, capable of in-order instruction execution, each computation entails driving all crossbar arrays within the core to enhance computational parallelism. The computational schematic is illustrated in Fig.~\ref{Fig::Model-1}(a), resulting in the compilation outcome depicted in Fig.~\ref{Fig::Model-1}(c), where inputs for each array are loaded, followed by simultaneous computation across all crossbar arrays. In contrast, accelerators featuring out-of-order instruction execution can operate as depicted in Fig.~\ref{Fig::Model-1}(b). Consequently, the compilation outcome, as shown in Fig.~\ref{Fig::Model-1}(d), achieves overlapping of memory access and computation.

\begin{figure}[t]
  \centering
  \includegraphics[width=0.85\linewidth]{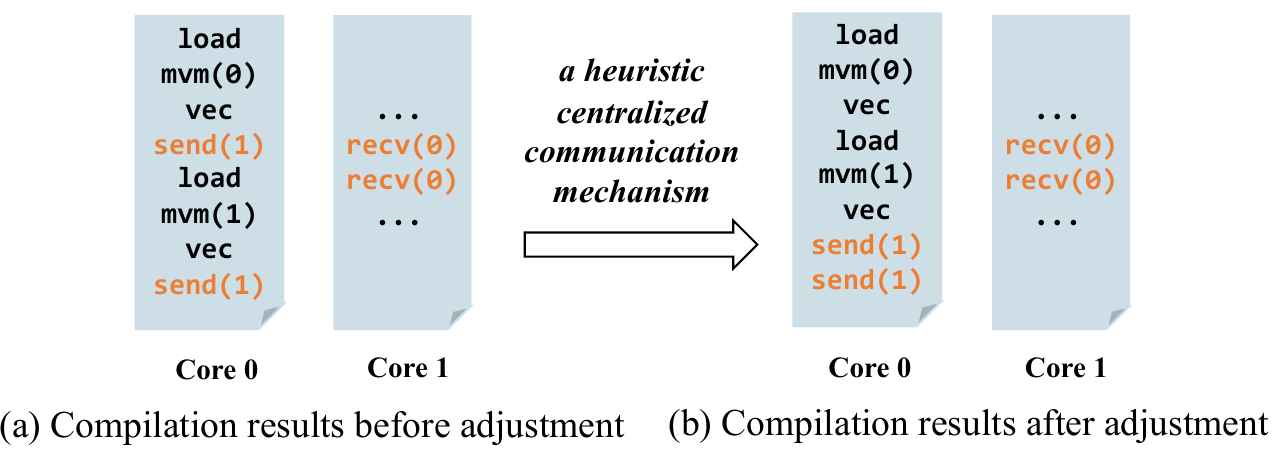}\\
  \caption{Example of adjustment based on communication mechanism. The arguments of \texttt{mvm}, \texttt{send}, and \texttt{recv} are the value of \texttt{\%idx}.}
  \label{Fig::Model-2}
\end{figure}

\textbf{(2) Communication mechanism}: For accelerators employing an asynchronous communication mechanism, typically only a \texttt{send} instruction is required, with \texttt{recv} implicitly completed by hardware interrupts or other signaling mechanisms. \emph{PIMCOMP} needs to record the address and length allocated by the compiler for data reception. For accelerators that adopt a synchronous communication mechanism, the compiler ensures that \texttt{send} and \texttt{recv} are paired to maintain the legitimacy of each inter-core communication. Since blocking occurs with each communication, we reduce synchronization overhead and enhance system parallelism by a heuristic centralized communication mechanism, adjusting the order of \texttt{send} and \texttt{recv}. For illustration, Fig.~\ref{Fig::Model-2}(a) shows the original compilation result, where the core waits for the completion of the first \texttt{send(1)} instruction, thereby hindering other crossbar arrays from performing MVM operations. Therefore, in Fig.~\ref{Fig::Model-2}(b), we adjust the position of the \texttt{send(1)} instruction to ensure the independent crossbar arrays complete the computational tasks in parallel, followed by the data transmission.

\begin{figure*}[t]
  \centering
  \includegraphics[width=0.7\linewidth]{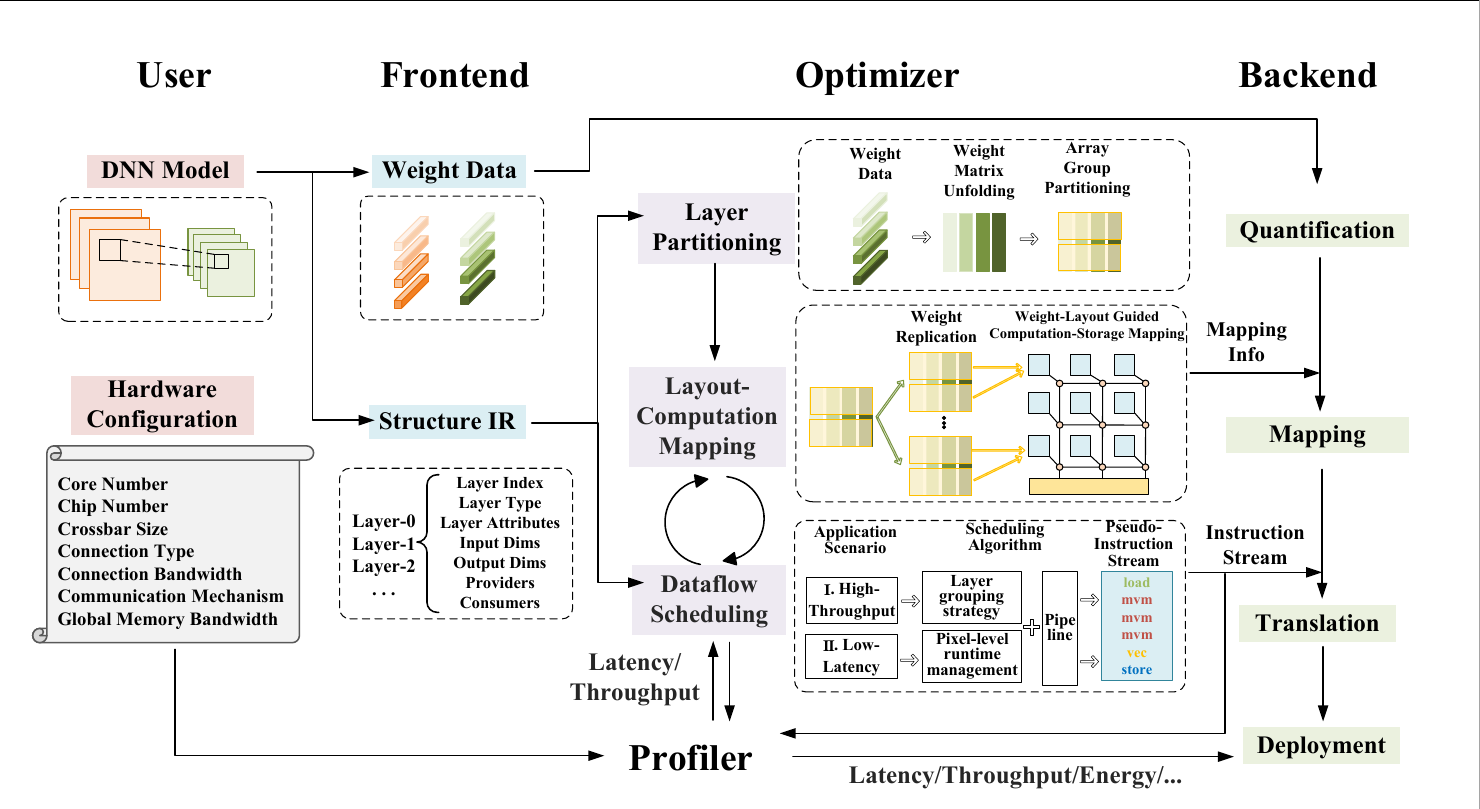}\\
  \caption{Overview of PIMCOMP.}
  \label{Fig::Overview}
\end{figure*}

\section{PIMCOMP Overview}
\label{Sec::Overview}

Fig.~\ref{Fig::Overview} illustrates the overview of \emph{PIMCOMP}, which consists of four main components: a frontend, an optimizer, a backend, and a profiler. The user inputs the DNN model and the hardware configuration, and \emph{PIMCOMP} automatically performs the deployment task. The \textbf{frontend} reads the DNN model and extracts the weight data and the structure intermediate representation (IR). Simultaneously, the frontend configures the optimizer and profiler based on the hardware configuration. In the \textbf{optimizer} stage, PIM-specific strategies are employed at multiple levels to enhance resource utilization and perform dataflow scheduling, producing weight mapping information and pseudo-instruction streams. In the \textbf{backend} stage, the weights are quantized and mapped according to the compiler outcomes, and the pseudo-instruction streams are translated into the actual hardware operation streams using the hardware library provided by the user. The \textbf{profiler} provides the performance metrics for the iterative optimization of the optimizer.

\subsection{Frontend}

The frontend parses the DNN model and converts it into IRs, eliminating the need for users to rewrite the model. \emph{PIMCOMP} supports DNN models described in the ONNX format, which can be easily obtained from various frameworks such as PyTorch and TensorFlow, thus enabling \emph{PIMCOMP}'s compatibility with multiple deep learning frameworks.

To enable efficient deployment, the frontend adopts a structure-weight splitting strategy, which parses the DNN model and produces separated structure IR and weight data. The structure IR contains the topology of the network and the parameters of the layers, which act as a lightweight proxy for the DNN model, facilitating fast iterative optimization in the optimizer stage. Moreover, the frontend performs graph-level optimizations such as layer fusion and elimination. The weight data consists of the actual weight values of each layer, which will be mapped to the physical hardware in the backend stage. Additionally, the frontend initializes the compiler and profiler based on the hardware configuration information.

\subsection{Optimizer}

The optimizer is the core component of  \emph{PIMCOMP}. We adopt a multi-level compilation approach focusing on three general stages to address the challenges associated with resource utilization and dataflow scheduling. Layer partitioning (Section~\ref{Sec::Partitioning}) specifies the compilation granularity for the entire compilation after unfolding the weight data into the matrix. Layout-computation mapping (Section~\ref{Sec::Mapping}) delineates the layout of weight data and the allocation of computational tasks among the PIM arrays considering weight replication. Dataflow scheduling (Section~\ref{Sec::Dataflow}) schedules the dataflow and constructs a pseudo-instruction stream for each core. The latter two stages form a closed-loop iterative optimization to boost inference performance under the performance feedback by the profiler.

\begin{figure}[t]
  \centering
  \includegraphics[width=1.0\linewidth]{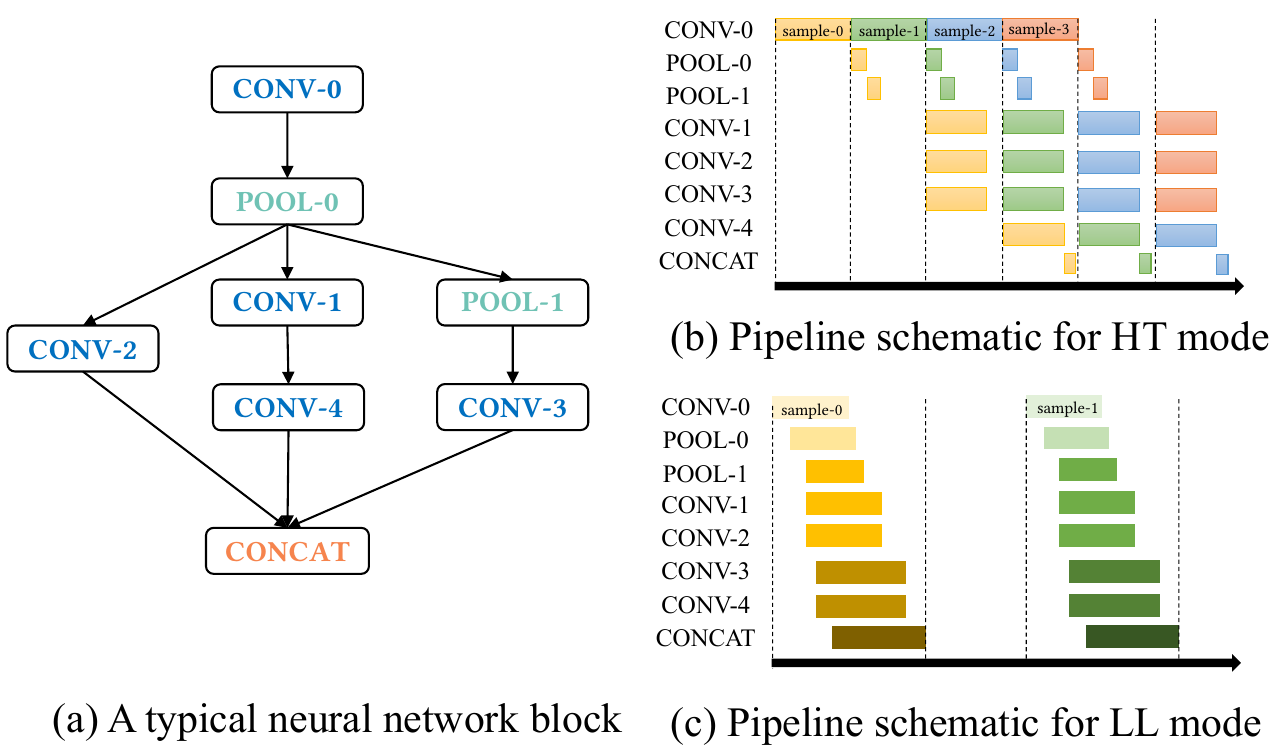}\\
  \caption{Pipeline schematic under HT and LL modes.}
  \label{Fig::Pipeline}
\end{figure}

To enhance the versatility of the compiler, \emph{PIMCOMP} has devised two compilation modes characterized by inter-layer pipelining: 
high-throughput (HT) and low-latency (LL), as detailed in Fig.~\ref{Fig::Pipeline}. The fundamental difference between them lies in the granularity of the inter-layer pipeline. In the HT mode, the DNN infers layer by layer, with different layers processing data from different samples. Inter-layer data transmission occurs after the computation of the current layer is complete, allowing for a higher degree of parallelism in computation execution, as computation is infrequently interrupted by communication. In the LL mode, as soon as a layer calculates an output, it passes it to the subsequent layer. As long as a layer receives enough data to perform computations, it can calculate promptly, thereby reducing inference latency.

\subsection{Backend}

The backend stage is responsible for interfacing \emph{PIMCOMP} with the actual hardware. As the  architecture abstraction obscures some specific hardware details, the backend stage refines and supplements them. Initially, the backend quantifies the weight data for each layer. Subsequently, it partitions the weight data, performs bit splitting, and then programs it into the actual PIM arrays based on the weight mapping information obtained from the optimizer. Thereafter, the backend translates the pseudo-instruction stream into hardware primitives, leveraging the hardware library provided by the user, which enables the physical accelerator to autonomously execute deployment tasks.
\emph{PIMCOMP} leaves an interface for the backend so that users can provide the backend for the specific hardware.

\subsection{Profiler}


The profiler provides performance metrics for the optimizer stage and pre-deployment evaluation for the accelerator. The profiler utilizes the architectural parameters provided by the user in Table~\ref{Tab::Parameter} and the pseudo-instruction stream to conduct performance and functional simulations of the  accelerator. In terms of latency, by establishing timing axes for multiple components, the profiler can simulate intra-core instruction-level parallel execution scenarios, taking into account the structural and data conflicts, as well as the synchronization overhead of inter-core communication. Additionally, if the user supplies power data for each component, the profiler can synthesize pseudo-instructions, establish the relationship between pseudo-instructions and components, and obtain the inference power dissipation. The profiler also collects statistical data such as resource utilization, memory footprint, inter-core data transmission, and memory access volume to inquire into the processing details.

\section{Layer Partitioning}
\label{Sec::Partitioning}

Facing accelerators containing thousands of PIM arrays, organizing and managing the numerous resources becomes a critical challenge for the programming model. Due to hardware array size limitations, the weight data of a layer is often distributed across multiple crossbar arrays. If the compiler uniformly manages all arrays within a layer, deployment flexibility is compromised. In cases where a core's PIM resources are insufficient to accommodate a layer, deployment may fail.
\textcolor{black}{On the other hand, if the compiler handles each array separately, the management flexibility is high. However, due to the exponential growth of the solution space, the toolchain's execution efficiency would be compromised, resulting in greater runtime overhead.}
To strike a balance between flexibility and efficiency, we propose the concept of \textit{array groups} as the fundamental unit within the programming model. In this section, we delve into the division rules and advantages of array groups. Additionally, we introduce the flexible unfolding format, which leverages different transformation methods to unfold the weight data.

\subsection{Array Group}
\label{Sec::Partitioning-1}

\begin{figure}[tbp]
  \centering
  \includegraphics[width=0.95\linewidth]{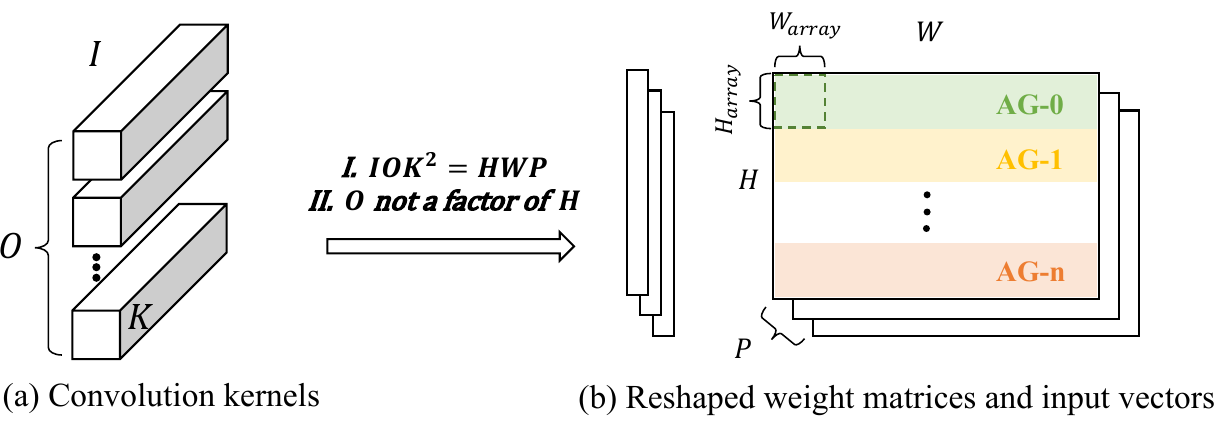}\\
  \caption{Unfolding strategy for convolutional kernels and partitioning strategy for the set of reshaped weight matrix.}
  \label{Fig::Partition}
\end{figure}

Given the ability of PIM arrays to perform MVM operations in constant time, a common practice for accelerating DNNs on PIM involves unfolding the weight data of a convolutional layer into a set of parallel weight matrices, unfolding the sliding window of input feature maps into a set of input vectors, and facilitating PIM arrays to compute the multiplication of weight matrices and input vectors. Suppose we have a convolutional layer whose weight data is unfolded (the specific unfolding process is described in Section~\ref{Sec::Partitioning-2}) into a set of $P$ weight matrices of size $H \times W$ each, and every sliding window is reshaped into $P$ input vectors of length $H$ each, as illustrated in Fig.~\ref{Fig::Partition}(b). We partition every weight matrix vertically into several groups based on the height of the array $H_{array}$ where each group, termed an array group (AG), contains $\lceil W/W_{array} \rceil$ arrays. These resulting array groups serve as the fundamental units for mapping and scheduling in the programming model of \emph{PIMCOMP}.

The rationales behind our design are as follows. From a hardware perspective, an AG physically corresponds to a set of crossbar arrays in a core. Since crossbar arrays within an AG share identical inputs, input data can be broadcasted to these arrays, thus avoiding redundant data transmissions and alleviating on-chip bandwidth. In terms of mapping flexibility, the existence of AGs decouples the rigid association between cores and layers. AGs from the same layer can be mapped to different cores, and a single core can accommodate AGs from distinct layers, allowing for flexible mapping of weight data and empowering compilers to explore a wide range of optimization possibilities. Regarding scheduling efficiency, arrays within the same AG can be managed by the same pseudo-instruction (\texttt{mvm}), enabling shared inputs and parallel outputs, thereby facilitating dataflow scheduling and pseudo-instruction generation. During the backend phase, pseudo-instructions at the granularity of AGs are transformed into primitives that match the accelerators' management granularity.

\subsection{Flexible Unfolding Format}
\label{Sec::Partitioning-2}

Section~\ref{Sec::Partitioning-1} highlights that utilizing PIM arrays to accelerate DNNs necessitates reshaping weight data into a set of 2D weight matrices. It is observed that different unfolding strategies may exhibit distinct computational and memory access characteristics. Consequently, the compiler needs to support flexible unfolding formats to address various hardware resource constraints and performance requirements. Hereafter, we exemplify the characteristics of different unfolding methods using a convolutional layer with input feature map size $F_{in}$, output feature map size $F_{out}$, input channel number $I$, output channel number $O$, and convolutional kernel size $K$.

As illustrated in Fig.~\ref{Fig::Partition}, each unfolding method can be represented by a tuple $(H,W,P)$, where $H$ and $W$ denote the height and width of the weight matrix, respectively, and $P$ represents the number of parallel weight matrices. Since the number of weight elements remains constant before and after unfolding, it follows that $HWP=IOK^2$. Additionally, as the direction of $H$ requires data accumulation, we can not choose $H$ such that $O$ is a factor of $H$ according to the convolution computational rule. Any unfolding method that satisfies these two rules is considered valid. The most common unfolding method is $(IK^2,O,1)$~\cite{MNSIM2,Pipelayer}, which unfolds the weight into a single matrix. To reuse input data and reduce loading overhead, $(I,O,K^2)$ is adopted in~\cite{Mapping}, requiring further accumulation of results from $K^2$ parallel weight matrices to obtain the complete result. Similarly, $(IK,OK,1)$ is utilized to achieve the same objective~\cite{CONV}. Furthermore, a reconfigurable architecture is introduced in~\cite{Benchmark} to support different unfolding strategies.

\begin{table}[t]
\caption{Theoretical performance of different weight unfolding methods.}
\label{Tab::Unfolding}
\centering
\scriptsize
\setlength{\tabcolsep}{1pt}
\begin{tabular}{|c|c|c|c|}
\hline
Type (H,W,P) & Computation cycle   & Load volume                    & Additional memory                     \\ \hline
$(I,O,K^2)$  & $F_{out}^2$ (\textcolor[RGB]{34,139,34}{S})     & $F_{in} F_{out} K I$ (\textcolor[RGB]{255,193,37}{M}) & $K^2I+K^2O$ (\textcolor[RGB]{202,12,22}{L})            \\ \hline
$(I,OK^2,1)$ & $F_{in}^2$ (\textcolor[RGB]{202,12,22}{L})      & $F_{in}^2 I$ (\textcolor[RGB]{34,139,34}{S})         & $I+K^2O$ (\textcolor[RGB]{255,193,37}{M}) \\ \hline
$(IK,O,K)$   & $F_{out}^2$ (\textcolor[RGB]{34,139,34}{S})     & $F_{in} F_{out} K I$ (\textcolor[RGB]{255,193,37}{M}) & $K^2 I + KO$ (\textcolor[RGB]{255,193,37}{M})           \\ \hline
$(IK,OK,1)$  & $F_{in}F_{out}$ (\textcolor[RGB]{255,193,37}{M}) & $F_{in} F_{out} K I$ (\textcolor[RGB]{255,193,37}{M}) & $KI+KO$ (\textcolor[RGB]{34,139,34}{S})                 \\ \hline
$(IK^2,O,1)$ & $F_{out}^2$ (\textcolor[RGB]{34,139,34}{S})     & $F_{out}^2 K^2 I$ (\textcolor[RGB]{202,12,22}{L})    & $K^2I+O$ (\textcolor[RGB]{255,193,37}{M})                    \\ \hline
\end{tabular}
\end{table}

Considering resource utilization, we have selected five valid unfolding methods, as shown in Table~\ref{Tab::Unfolding}. Besides, we outline the theoretical computation cycles, data loading volume, and additional memory requirements. Here, the data loading volume refers to the amount of input data retrieved from the global memory, while the additional memory requirement denotes the memory overhead needed for input reuse or intermediate result storage. Users can flexibly select the unfolding method based on hardware resources and optimization objectives. In this work, we employ a greedy method to determine the unfolding scheme: prioritizing minimal computation cycles while pursuing minimal data loading volume in the HT mode and minimal memory requirement in the LL mode.

\section{Layout-Computation Mapping}
\label{Sec::Mapping}

During the DNN inference process, each AG undergoes numerous iterations of load-computation-transmission, which requires the comprehensive utilization of hardware computation, memory, and communication capabilities. The spatial distribution of AGs across the cores and the allocation of computational tasks directly influence these three system characteristics. Furthermore, weight replication is a crucial yet often overlooked aspect by previous works, as it enhances resource utilization and improves performance.

To adhere to the characteristics of PIM and fully exploit hardware performance, we propose the weight-layout guided computation-storage-mapping. For weight layout, the genetic algorithm (GA) is utilized to determine the weight replication and distribution of AGs throughout all the cores (Section~\ref{Sec::Mapping-1}). For computation-storage-mapping, we flexibly determine the allocation of computational tasks to AGs based on the weight layout for both pipeline modes (Section~\ref{Sec::Mapping-2}).

\subsection{GA-based Weight Layout Optimization}
\label{Sec::Mapping-1}

\begin{figure}[tp]
  \centering
  \includegraphics[width=0.80\linewidth]{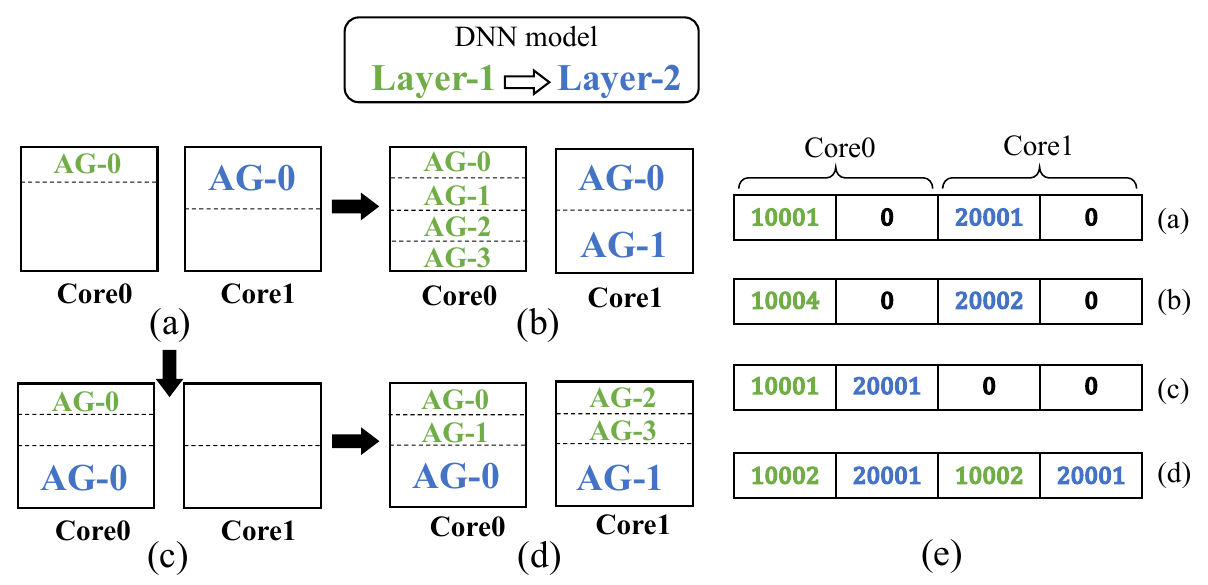}\\
  \caption{Example of using GA to optimize weight layout considering weight replication. (a) Common practice. (b) Preferred approach of HT mode. (c) Intermediate result. (d) Preferred approach of LL mode. (e) Chromosome of (a)-(d).}
  \label{Fig::Mapping}
\end{figure}

As the number of replicas in a layer increases, the amount of AGs it contains also rises, thus expanding the search space for AG layout. To explore the vast solution space fully, we propose employing GA to jointly optimize weight replication and AG layout. To represent the mapping relationship between all cores and AGs using a single chromosome, we adopt a value-position combined encoding method, where each gene's value represents the number of AGs, and its position signifies the mapped core. Specifically, we define the gene length for each core as $max\_layer\_num\_in\_core$, constraining the number of layers each core can accommodate. Consequently, the length of each chromosome is $max\_layer\_num\_in\_core \times core\_num$. The numerical value of each gene is an integer, calculated as $layer\_index \times 10000 + AG\_num$ denoting  $AG\_num$ AGs of the layer with index $layer\_index$, with $10000$ indicating the core can accommodate up to $9999$ AGs for that layer. Fig.~\ref{Fig::Mapping}(e) illustrates the chromosomes for mappings Figs.~\ref{Fig::Mapping}(a)-(d) when $max\_layer\_num\_in\_core=2$.

In the initialization phase, we set the replication factor of each layer to 1. Mutation is a critical phase for performance enhancement. We devise the following mutation strategies to explore the solution space composed of AG distribution and weight replication: \uppercase\expandafter{\romannumeral1}. Randomly select a gene and either increase or decrease its AG count. If the gene's value is 0 originally, indicating an unassigned position, we randomly assign a layer to it with a replication factor of 1. \uppercase\expandafter{\romannumeral2}. Randomly select two genes and partially swap their positions. Here, ``partially'' implies that each gene may only contribute a portion of AGs for exchange, and ``swap'' denotes placing the exchanged AGs into the zero-valued positions of the original gene's core. We ensure the legality of each chromosome throughout the mutation process. For the fitness function, our previous work~\cite{PIMCOMP} employs analytical models for evaluation.
In this work, we assess using the profiler. Specifically, for each mapping represented by a chromosome, we perform the dataflow scheduling process mentioned later to generate a pseudo-instruction stream. This stream is not the final complete sequence; rather, it is a simplified version used for rapid evaluation by the profiler to obtain the performance considering computation, memory access, and communication as the fitness function.

Fig.~\ref{Fig::Mapping} presents an example of using GA to optimize the weight layout for the HT and LL modes, where the PIM system includes two cores, and the DNN comprises two convolutional layers. Each layer has one AG, with the AG from Layer-1 and Layer-2 occupying 1/4 and 1/2 of the PIM resources within a core, respectively. Fig.~\ref{Fig::Mapping}(a) depicts the original method, where each core contains only one layer\textemdash a common approach in previous works~\cite{ISAAC, Pipelayer, MNSIM2}. This leads to suboptimal resource utilization and failure to leverage PIM's parallel computing advantages. For the HT mode, with modest inter-core communication, the key to improve performance lies in reducing memory access. The GA tends to find the mapping shown in Fig.~\ref{Fig::Mapping}(b), where Layer-1 and Layer-2 have replication factors of 4 and 2, respectively, with their AGs clustered. This approach maximizes resource utilization and enhances data reuse potential due to overlapping sliding windows of convolutional layers. For the LL mode, communication often becomes the bottleneck; hence, achieving a balance between computation and communication is the optimization direction. The GA tends to find the mapping shown in Fig.~\ref{Fig::Mapping}(d), where data produced by Layer-1 can directly serve Layer-2 within the same core, thereby reducing a portion of data communication.

\begin{figure}[t]
  \centering
  \includegraphics[width=0.95\linewidth]{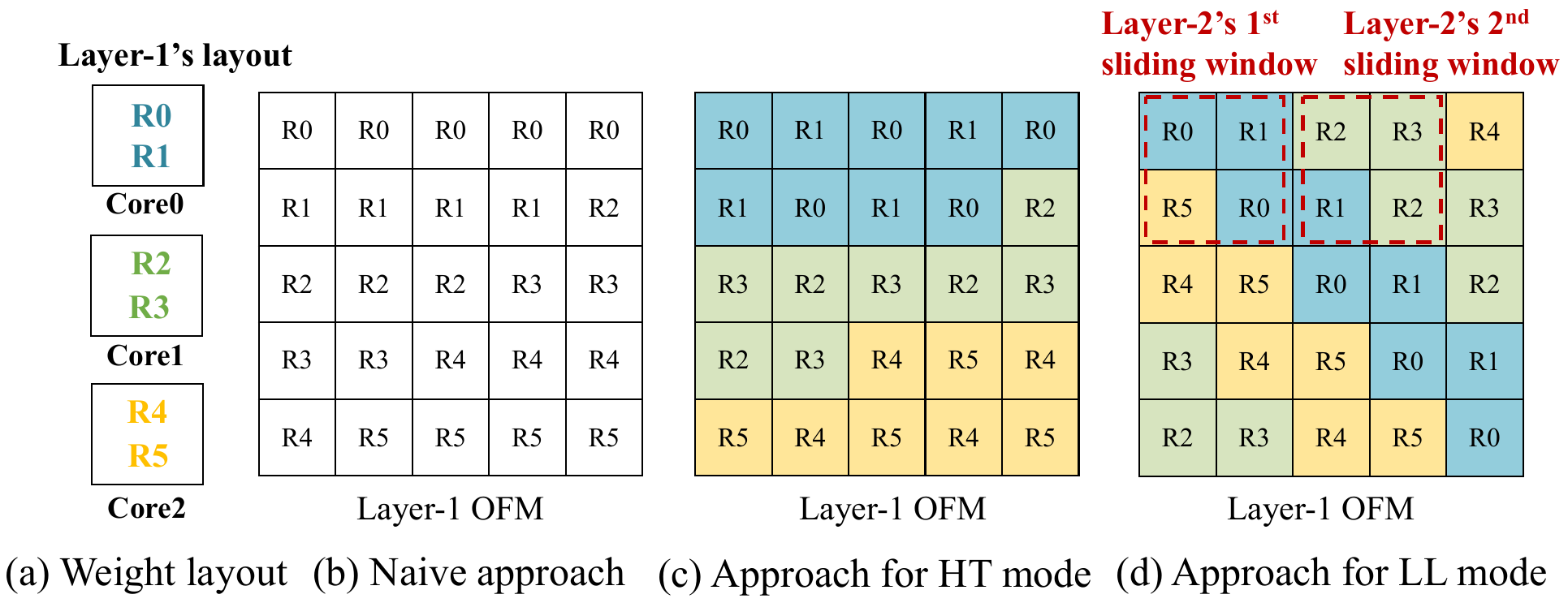}\\
  \caption{Different computational tasks allocation approaches.}
  \label{Fig::Task}
\end{figure}

\subsection{Adaptive Computation-Storage-Mapping}
\label{Sec::Mapping-2}

A convolutional layer exhibits local connectivity, thus its inference process involves multiple convolution operators, each performing computation between a sliding window and weight data. According to the computational features of PIM and the unfolding method employed by \emph{PIMCOMP}, the computational task of each convolution operator is allocated to a replica of the convolutional layer. Fig.~\ref{Fig::Task} illustrates an example of computational task allocation by \emph{PIMCOMP}. Layer-1 possesses 6 replicas (denoted as R0 to R5 in Fig.~\ref{Fig::Task}(a)), each housing one AG, evenly distributed across Core-0 to Core-2. For illustrative purposes, in Figs.~\ref{Fig::Task}(b)(c)(d), we represent the computational task allocation for Layer-1 as the partitioning of the output feature map (OFM), where each point in the OFM represents the computational task of one convolution operator. Fig.~\ref{Fig::Task}(b) depicts the original method of evenly distributing computational tasks among the various replicas.

Fig.~\ref{Fig::Task}(c) shows the allocation approach in the HT mode. For each layer, the number of computational tasks undertaken by each core is proportional to the number of replicas of that layer housed within the core, ensuring a uniform distribution of computational tasks across all cores. Due to the characteristics of convolutional layers, there is typically an overlap between adjacent sliding windows in the input data, which is pre-prepared in the HT mode. Consequently, each replica alternately completes adjacent computational tasks within each core, enabling input data reuse and reducing global memory access. Although allocation in Fig.~\ref{Fig::Task}(b) also achieves input reuse, when facing larger input feature maps, there is no overlap between the input windows of R0 and R1. Therefore, R0 and R1 need to separately maintain a portion of memory for data reuse. In contrast, in Fig.~\ref{Fig::Task}(c), R0 and R1 can share input data, significantly reducing memory requirements.

Fig.~\ref{Fig::Task}(d) depicts the allocation approach in the LL mode. 
To rapidly generate results for subsequent layers, we utilize complete computational parallelism to execute computational tasks sequentially. This allocation approach will increase data communication to some extent. Therefore, in Section~\ref{Sec::Dataflow}, we will introduce a pixel-level transmission and memory allocation mechanism designed for the LL mode to reduce communication overhead while reducing memory requirements. For non-convolutional and non-fully-connected layers, which utilize the VFUs within the core, our allocation criterion is to minimize data communication greedily. As indicated by the red dashed box in Fig.~\ref{Fig::Task}(d), assuming Layer-2 is a pooling layer, its first windowed operation is completed in Core-0 since three-quarters of the data needed by the first window is stored in Core-0, and the second windowed operation is completed in Core-1 likewise. Consequently, we replicate Layer-2 and distribute the computational tasks among multiple cores according to this rule.

\section{Dataflow Scheduling}
\label{Sec::Dataflow}

In general, PIM accelerators store the weights in PIM crossbar arrays, which also serve as computational units. Ideally, these arrays are capable of operating in parallel.
The nature of PIM endow DNN models with the potential for concurrent computation across various layers. Consequently, this leads to the development of an on-chip pipeline suitable for PIM, enabling inter-layer network pipelining on the chip, without the need for multiple chips. To cater to different application scenarios, we design the HT and LL modes by adjusting the granularity of the pipeline. This section presents the optimization considerations for each mode and the corresponding scheduling algorithms that generate pseudo-instruction streams for each core based on the allocation result of computational tasks obtained in Section~\ref{Sec::Mapping}.

\subsection{High-Throughput Mode}

In the HT mode, a DNN infers layer by layer, suitable for scenarios involving continuous input streams. Fig.~\ref{Fig::Pipeline}(b) illustrates its primary characteristics: layers with data dependencies do not overlap in runtime within the same cycle. Consequently, each layer preserves its output via the global or local memory, transmitting it to its consumers upon the cycle's completion.

However, in scenarios like autonomous driving that demand high-throughput and real-time capabilities, the first-batch latency directly impacts decision speed and response time. To enhance system service quality, we propose a layer grouping strategy optimized for first-batch latency. It groups the layers of the original computational graph based on two principles: \uppercase\expandafter{\romannumeral1}. Group data-dependent layers without compromising throughput, and \uppercase\expandafter{\romannumeral2}. Group independent layers for concurrent execution. During scheduling, layer groups are executed in the topological order. Independent layers within a group can execute simultaneously, while data-dependent layers follow the topological sequence. These rules reduce the first-batch latency from the product of the number of layers and the cycle to the product of the number of layer groups and the cycle.

Taking Fig.~\ref{Fig::Pipeline}(b) with 8 layers as an example, assume the time $T_{CONV0}$ required to execute layer CONV-0 is the longest, dictating the pipeline's cycle. The first-batch latency would normally be $8 T_{CONV0}$. We observe that the combined execution time of POOL-0 and POOL-1 is less than $T_{CONV0}$, allowing two pooling layers to complete within the same cycle without worsening the pipeline cycle, so we group POOL-0 and POOL-1 together. Similarly, we group CONV-4 and CONCAT layers together. Subsequently, CONV1-3 are three independent layers, thus grouped to compute simultaneously within one cycle. The group strategy results in 4 layer groups, offering a 50\% reduction in the first-batch latency metric.

\begin{algorithm}[t]
\small
\caption{High-throughput dataflow scheduling.}
\label{Alg::HT}
\KwIn{LayerGroupList}

CompletedLayerList $\leftarrow$ []  \\  
\While{CompletedLayerList.Len() != LayerNum}
{
    \ForEach{LayerGroup in LayerGroupList}
    {
         IndepLayerList $\leftarrow$ LayerGroup.Layers.IndependentLayers() \\
        \ForEach{Layer in IndepLayerList}
        {
            \uIf{Layer.Op == CONV or FC}
            {
                \ForEach{RepLayer in Layer.ReplicationList}
                {
                    \If{RepLayer.TaskIdx $<$ RepLayer.TaskNum}
                    {
                        RepLayer.Run(RepLayer.TaskIdx) \\
                        RepLayer.TaskIdx += 1  \\  
                        Layer.TaskIdx += 1  \\  
                    }
                }
            }
            \Else
            {
                Layer.Run(Layer.TaskIdx, CoreNum)  \\
                Layer.TaskIdx += CoreNum  \\    
            }
            \If{Layer.TaskIdx == Layer.TaskNum}
            {
                CompletedLayerList.Append(Layer.Idx)  \\  
            }
        }
        LayerGroup.Layers $\leftarrow$ UpdateLayers(LayerGroup.Layers, CompletedLayerList) \\  
    }
}

\end{algorithm}

Algorithm~\ref{Alg::HT} delineates the convolution operator-level HT dataflow scheduling algorithm. Since the hardware exhibits consistent behavior across different batches, the accelerator can reuse the same pseudo-instructions, so Algorithm~\ref{Alg::HT} initially generates a saturated state pipeline throughout all layer groups, followed by the backend stage controlling the pipeline's filling and emptying processes. Lines 2-5 describe the scheduling strategy, where \texttt{LayerGroupList} is a list that arranges the layer groups obtained by the layer grouping strategy based on the topological order. 
In line 9, \texttt{RepLayer.Run} signifies the launch of a computational task, which, within \emph{PIMCOMP}, is essentially a convolution operator, thus enabling fine-grained convolution operator-level scheduling to exploit hardware compute parallelism. The \texttt{Run} method includes reading one sliding window of the input feature map with consideration of data reuse, triggering the computation of each AG within the replica, and accumulating partial sums between AGs. If this layer is fused with an activation layer, the activation operation will be completed instantly using the VFUs. The result of this task, i.e., a new output pixel, will be stored back to global memory or retained on the chip based on the capacity of the on-chip local memory. Lines 12-14 indicates that layers beyond convolutional and fully-connected layers are also allocated to multiple cores to enhance computational parallelism. Line 17 entails the removal of layers that have completed computational tasks.

\subsection{Low-Latency Mode}

In the LL mode, to expedite the attainment of final results, each layer promptly transmits output results to subsequent layers via on-chip connections. This operational characteristic poses two challenges for compilation: \uppercase\expandafter{\romannumeral1}. On-chip communication overhead, and \uppercase\expandafter{\romannumeral2}. Runtime management of computational tasks and on-chip storage. We address the former by achieving a balance between computation and communication through layout-computation mapping, as detailed in Section~\ref{Sec::Mapping}. The latter challenge is unique to the fine-grained pipeline of PIM on-chip architecture.
In this mode, aside from the first layer's input being read from global memory, the input data for other layers come from the outputs of their predecessor layers during runtime. Consequently, each AG stores the received data in local memory. These data cannot be immediately utilized but need to wait until enough data is received to perform the corresponding computational task.
Due to the limited capacity of the on-chip local memory, the allocation of new memory block for received data cannot be indefinite; timely clearance of subsequently unused data is necessary to reclaim memory. Furthermore, in DNNs, a layer may have multiple predecessor and successor layers, each comprising multiple replicas and each replica composed of multiple AGs, thus compounding the complexity of this process.

\begin{figure}[tp]
  \centering
  \includegraphics[width=0.95\linewidth]{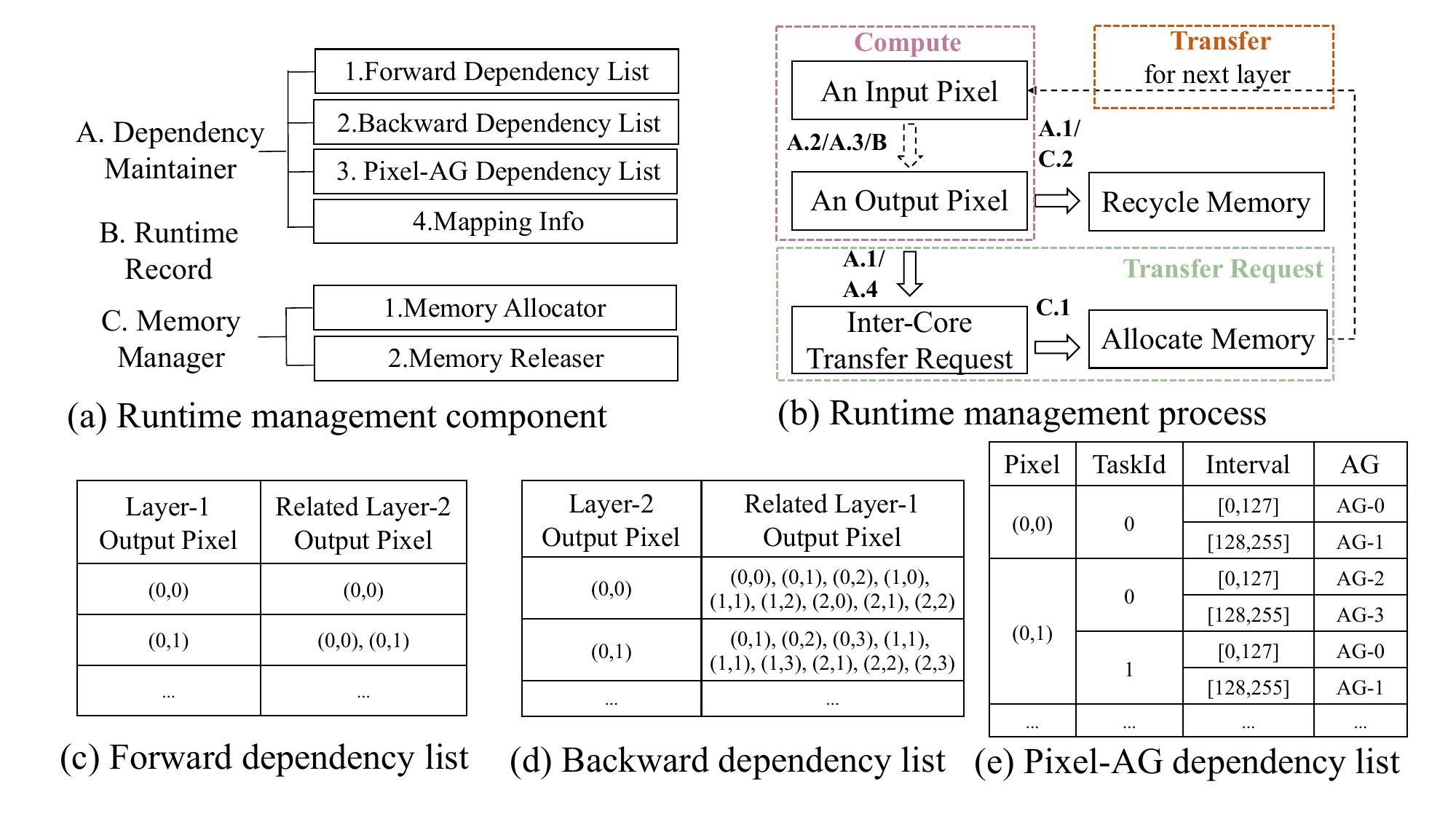}\\
  \caption{Pixel-level runtime management strategy. }
  \label{Fig::Memory}
\end{figure}

To efficiently control the computation process, optimize memory usage, and reduce data transmission overhead, we propose the pixel-level runtime management strategy as illustrated in Fig.~\ref{Fig::Memory}, \textcolor{black}{which serves as a virtual runtime mechanism in software}. Since each computational task outputs a complete pixel, we designate the pixel as the unit for memory allocation and data transmission. Fig.~\ref{Fig::Memory}(a) illustrates the components required for management. 
With data transmission occurring between cores, data dependencies existing between replicas of layers, and the granularity of transmission and allocation being the pixel, we establish multi-level dependencies through the dependency maintainer.
The forward dependency list records the relationship between a layer's output pixels and its successor layers' output pixels. Inversely, the backward dependency list documents the relationship between a layer's output pixels and its predecessor layers' output pixels. The pixel-AG dependency list records the correspondence between AGs and input pixels across different computational tasks. The mapping info is the mapping relationship between AGs and cores obtained in Section~\ref{Sec::Mapping}. Additionally, we design the runtime record to log the address of each pixel within each core. To achieve flexible memory allocation, \emph{PIMCOMP} adopts a heap-based memory manager for  allocation and release.

\begin{algorithm}[t]
    \small
    \caption{Low-latency dataflow scheduling.}
    \label{Alg::LL}
    \KwIn{LayerQueue}

    TransReqQueue $\leftarrow$ Queue() \\
    \While{not LayerQueue.Empty()}
    {
        Layer $\leftarrow$ LayerQueue.Pop()  \\
        
        \ForEach{RepLayer in Layer.ReplicationList}
        {
            \If{RepLayer.Ready(RepLayer.TaskIdx)}
            {
                Result $\leftarrow$ RepLayer.Compute(RepLayer.TaskIdx) \\
                TransReq $\leftarrow$ RepLayer.TransmitRequest(Result) \\
                TransReqQueue.Append(TransReq) \\
                RepLayer.TaskIdx += 1 \\
                Layer.TaskIdx += 1 \\
            }
        }
        \If{Layer.TaskIdx $<$ Layer.TaskNum}
        {
            LayerQueue.Append(Layer) \\
        }
        \If{Meet Transmit Conditions}
        {
          PerformTransmit(TransReqQueue) \\
        }
    }
\end{algorithm}

Fig.~\ref{Fig::Memory}(b) illustrates the management process. When a replica of a layer receives an input pixel, if \emph{PIMCOMP} detects the feasibility of performing a computational task, it controls the AGs of this replica to calculate. For the result output pixel, we generate inter-core transmission requests based on A.1 and A.4, ensuring no redundant transmission between any two cores. Subsequently, these requests invoke C.1 for allocating local memory addresses for each pixel in the receiving core. Finally, data transmissions are carried out at an appropriate time, with these transmitted data serving as inputs for successor layers, thus re-entering the process. Upon completion of each computational task, we utilize A.1 and C.2 to delete pixels that will no longer be used by any other replicas, thus reclaiming memory.

Based on this foundation, we propose the LL dataflow scheduling algorithm, as shown in Algorithm~\ref{Alg::LL}. We also adopt convolution operator-level scheduling. The input parameter \texttt{LayerQueue} is a queue of layers arranged in the topological order. 
In lines 5 and 6, \texttt{RepLayer.Ready} and \texttt{RepLayer.Compute} correspond to the Compute process in Fig.~\ref{Fig::Memory}(b). Unlike the \texttt{RepLayer.Run} method in the HT mode, the \texttt{Compute} method merely controls the computation process without engaging in data transmission. Regarding transmission requests, we do not execute them directly due to the reasons illustrated in Fig.~\ref{Fig::Model-2}, where frequent data transmissions can impact parallelism. Instead, we adopt a heuristic centralized communication mechanism: we cache data transmission requests first (lines 7 and 8) and initiate data transmission only when the number of transmission requests in the system reaches a certain threshold or when the last layer is traversed (lines 13 and 14). The request information for data transmission includes the data, the two cores involved in communication, and the pre-assigned receiving address. We ensure that memory is not reclaimed or overwritten before data transmission occurs.

\section{Evaluations}
\label{Sec::Evaluation}

\subsection{Experimental Setup}

\begin{table}[t]
\centering
\caption{Configuration of three existing architectures.}
\label{Tab::Hardware}
\setlength{\tabcolsep}{2pt}
\begin{tabular}{|c|c|c|c|c|c|c|}
\hline
       & \#chip & \#core & \#crossbar & \begin{tabular}[c]{@{}c@{}}Crossbar \\ array size\end{tabular} & \begin{tabular}[c]{@{}c@{}}Cell\\ precision\end{tabular} & Capacity \\ \hline
Arch-A~\cite{ISAAC} & 1      & 168    & 96                                                            & 128$\times$128                                                  & 2-bit                                                    & 63MB     \\ \hline
Arch-B~\cite{PUMA} & 1      & 138    & 128                                                            & 128$\times$128                                                  & 2-bit                                                    & 69MB     \\ \hline
Arch-C~\cite{ISSCC} & 16     & 4      & 8                                                               & 512$\times$1024                                                 & 2-bit                                                    & 64MB     \\ \hline
\end{tabular}
\end{table}

To validate the generality of \emph{PIMCOMP}, we instantiate the abstract hardware using three different architectures with the corresponding configurations presented in Table~\ref{Tab::Hardware} and additional parameters referenced from existing literature~\cite{ISAAC}. We expand the number of chips of \cite{ISSCC} to 16 to ensure sufficient PIM resources to accommodate complete DNN models and avoid weight rewriting. The power data for the crossbar arrays and VFU units come from \cite{PUMA}. The memory and router modules are simulated using CACTI~\cite{CACTI} and Orion 3.0~\cite{ORION} to obtain power data, which are then scaled to the 32nm technology node.

For the comparison of compilation methods, we select \emph{SongC}~\cite{SONGC}, \emph{PUMA}~\cite{PUMA}, and \emph{Polyhedral}~\cite{POLY} as baselines, whose characteristics are elaborated in Section~\ref{Sec::Tools}. We implement these three methods faithfully within our compiler. Our comparison focuses on the impact of differences in scheduling granularity, resource allocation, data layout, and dataflow scheduling on performance. Therefore, the optimization methods designed by \emph{PIMCOMP} for unfolding weight data, alleviating communication overhead, and reducing memory access are applied equally to each implementation. We conduct simulation through the profiler described in Section~\ref{Sec::Overview}, which provides performance metrics such as inference latency, throughput, and energy consumption, along with detailed runtime information, including resource utilization, global memory access volume, and local memory overhead.

Four common networks are selected as benchmarks: vgg8~\cite{VGG} and resnet18~\cite{Resnet} trained on the MNIST dataset, and resnet34~\cite{Resnet} and googlenet~\cite{Googlenet} trained on the ImageNet datasets. The weight data for these models are quantized to 16-bit fixed-point numbers. In the HT scenario, the batch size is set to 128, while in the LL mode, the batch size is 1.

\begin{figure}[t]
 \centering
        \includegraphics[width=0.98\linewidth]{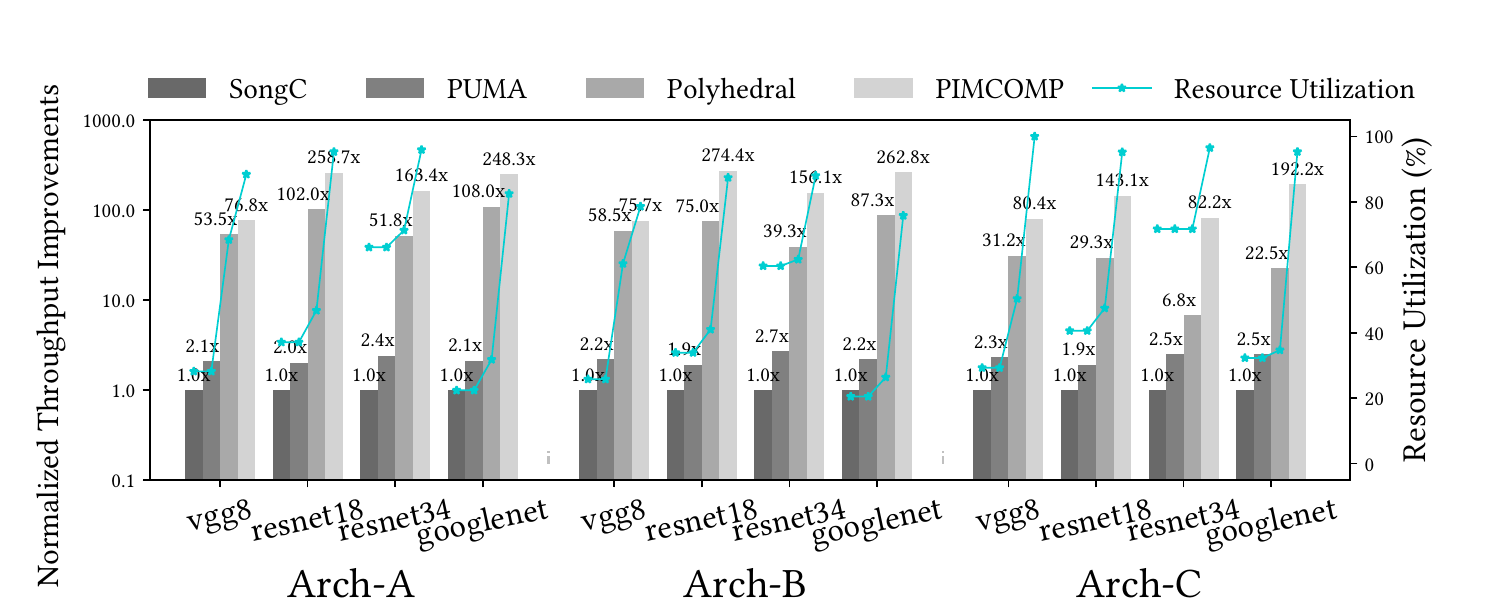}
        \caption{Throughput and resource utilization of HT scenario.}
        \label{Fig::HT-throughput}
\end{figure}

\begin{figure}[t]
\centering
        \includegraphics[width=0.98\linewidth]{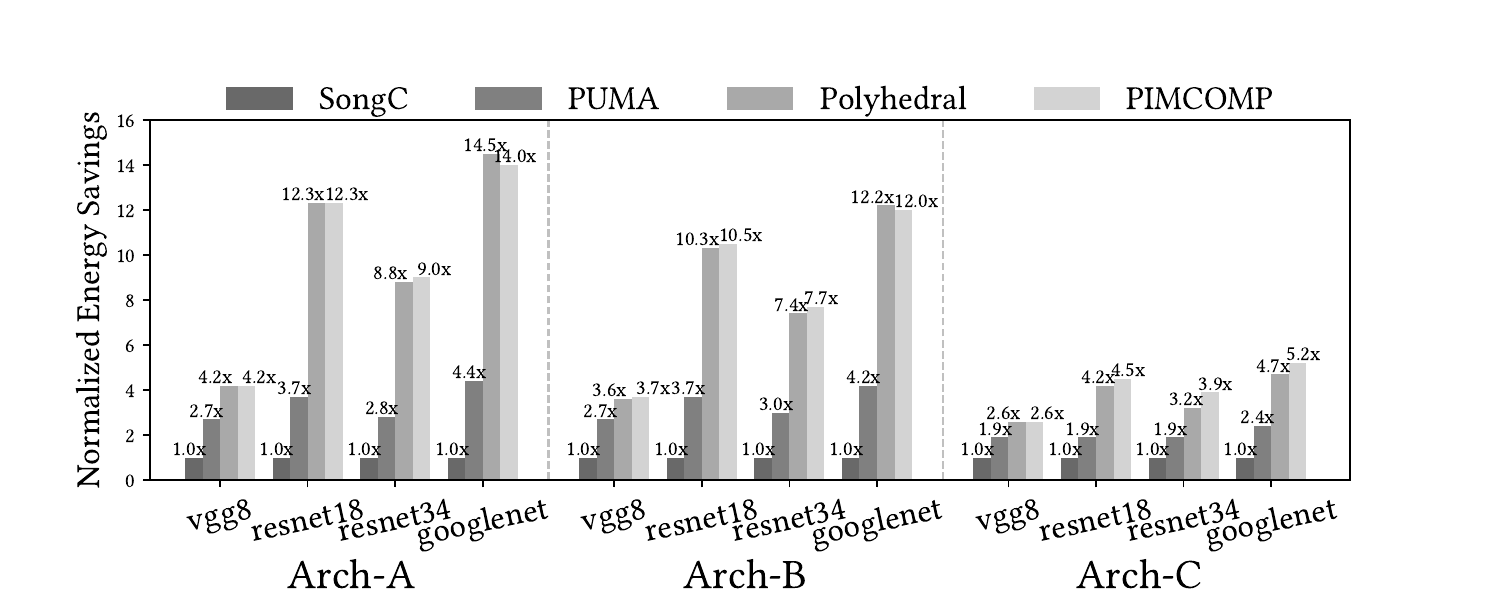}
        \caption{Energy savings of HT scenario.}
        \label{Fig::HT-energy}
\end{figure}

\begin{figure}[t]
\centering
        \includegraphics[width=0.98\linewidth]{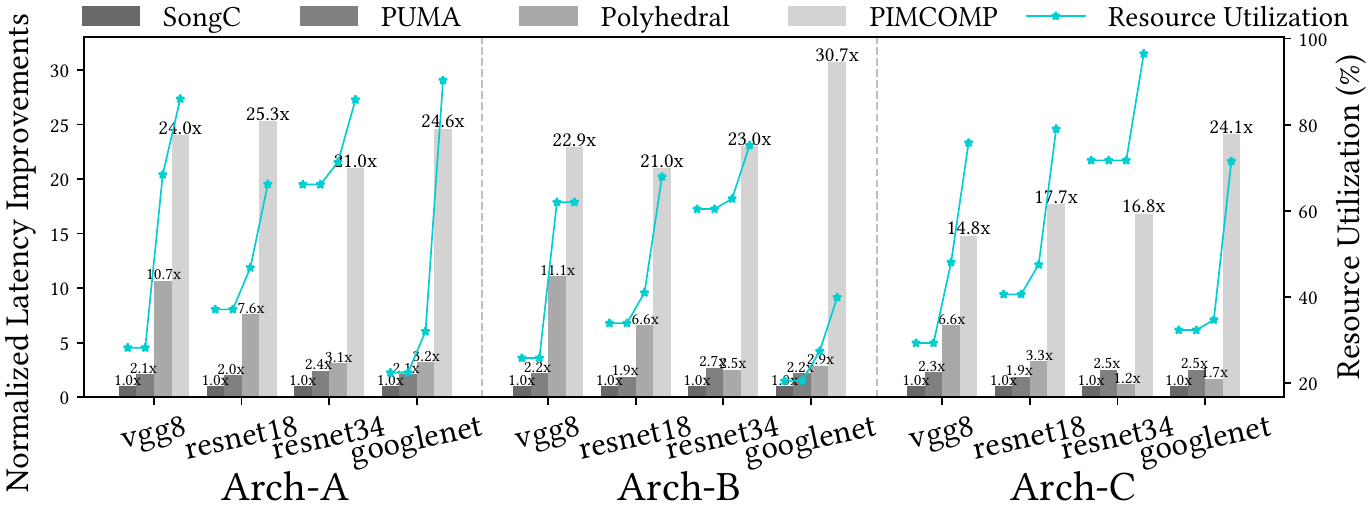}
        \caption{Latency and resource utilization of LL scenario.}
        \label{Fig::LL-latency}
\end{figure}

\begin{figure}[t]
\centering
        \includegraphics[width=0.98\linewidth]{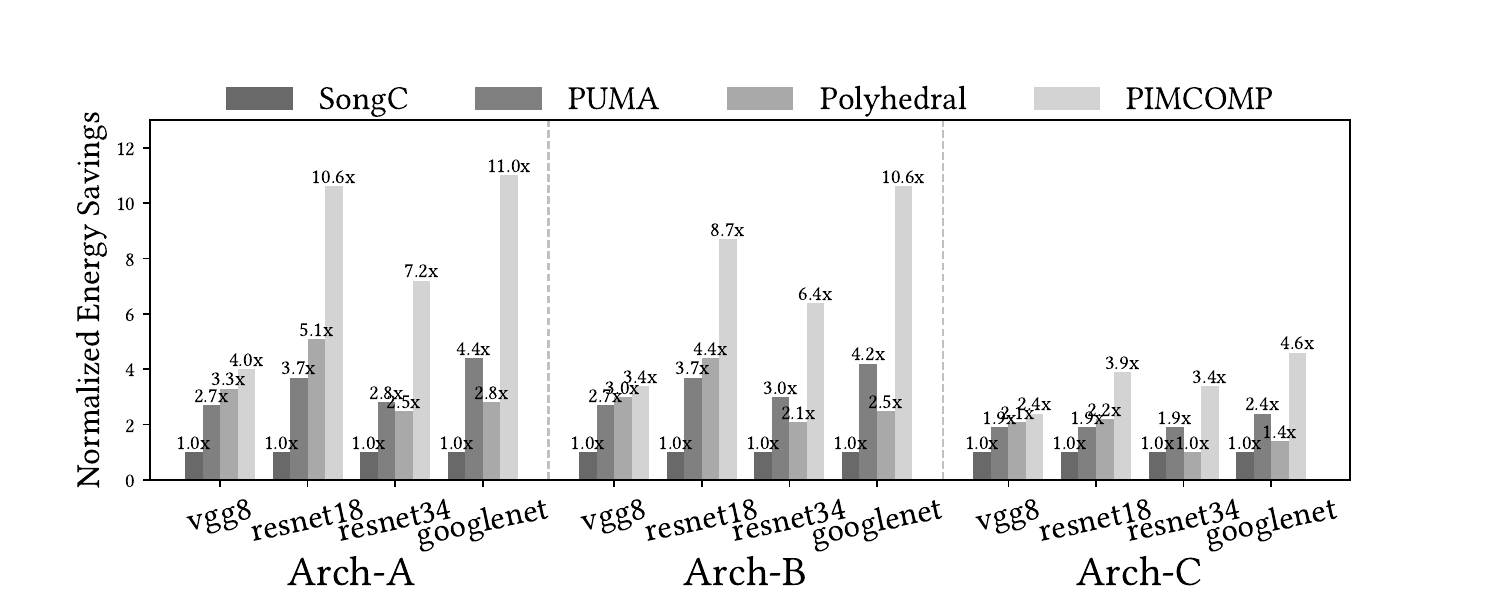}
        \caption{Energy savings of LL scenario.}
        \label{Fig::LL-energy}
\end{figure}

\subsection{End-to-End Inference Results}

\subsubsection{High Throughput Mode}

Fig.~\ref{Fig::HT-throughput} presents the throughput and resource utilization of different compilation strategies across three architectures in the HT scenario. \emph{PIMCOMP} achieves significant enhancement across all benchmarks on all architectures. Both \emph{Polyhedral}~\cite{POLY} and \emph{PIMCOMP} employ weight replication methods and sample-granularity pipelining to increase hardware computation parallelism, resulting in throughput improvements of 44.7$\times$ and 149.5$\times$ over \emph{SongC}~\cite{SONGC}, respectively. However, \emph{Polyhedral}~\cite{POLY}’s mapping and scheduling granularity is at the layer level, where each core can only accommodate one replica of a layer, leading to idle resources. Regarding resource utilization, \emph{PIMCOMP} defines the array group as the mapping granularity, enabling a finer layout mapping and allowing for a higher weight replication factor (with an average utilization improvement of 38.8\% over \emph{Polyhedral}~\cite{POLY}). In terms of scheduling granularity, \emph{PIMCOMP} initiates a sliding window computation task each time, ensuring that arrays without structural conflicts within the core can operate in parallel. Consequently, \emph{PIMCOMP} achieves an average throughput improvement of 3.3$\times$ over \emph{Polyhedral}~\cite{POLY}.

Fig.~\ref{Fig::HT-energy} compares the inference energy in the HT mode. \emph{PIMCOMP} achieves significant energy savings across all three architectures, primarily because \emph{PIMCOMP} boosts throughput, reducing total inference time and lowering the system's leakage energy. In the HT mode, the energy consumption of \emph{PIMCOMP} and \emph{Polyhedral}~\cite{POLY} are nearly identical because both significantly reduce the time to complete inference, making dynamic energy the dominant of total energy. Since the computational workload for inferring the same network remains constant across different compilation methods, \emph{PIMCOMP} and \emph{Polyhedral}~\cite{POLY} exhibit similar dynamic energy and total energy consumption. Overall, \emph{PIMCOMP} achieves an average improvement of 9.0$\times$ and 7.7$\times$ over \emph{SongC}~\cite{SONGC} on Arch-A and Arch-B, respectively, but only a 3.9$\times$ improvement on Arch-C. This is because Arch-C has fewer cores than Arch-A and Arch-B, resulting in lower system leakage power, while the larger crossbar arrays in Arch-C incur higher dynamic power. Consequently, as \emph{PIMCOMP} reduces energy consumption by lowering leakage energy, the optimization effect on Arch-C is somewhat less pronounced than on Arch-A and Arch-B.

\subsubsection{Low Latency Mode}

Fig.~\ref{Fig::LL-latency} illustrates the inference latency and resource utilization of different compilation strategies in the LL scenario. \emph{PIMCOMP} demonstrates a notable improvement over the other compilation methods on various architectures. Although \emph{PUMA}~\cite{PUMA} introduces inter-layer pipelining, it does not implement weight replication to utilize resources fully, resulting in imbalanced inter-layer execution times. The sample-granularity pipelining of \emph{Polyhedral}~\cite{POLY} fails to meet the low-latency requirements. \emph{PIMCOMP} designs inter-layer pipelines at the granularity of the convolution operator, achieving inference latency improvements of 21.8$\times$, 9.8$\times$, and 5.4$\times$ over \emph{SongC}~\cite{SONGC}, \emph{PUMA}~\cite{PUMA}, and \emph{Polyhedral}~\cite{POLY}, respectively, when the batch size is 1.

Fig.~\ref{Fig::LL-energy} displays the comparison of inference energy in the LL mode. \emph{PIMCOMP} significantly reduces the latency, resulting in energy savings of 5.6$\times$, 2.0$\times$, and 2.3$\times$ compared with \emph{SongC}\cite{SONGC}, \emph{PUMA}\cite{PUMA}, and \emph{Polyhedral}\cite{POLY}, respectively. The energy-saving effect in the LL mode is somewhat diminished compared with the HT mode, because the inference time per sample with a batch size of 1 is higher than the average time per sample with a batch size of 128, resulting in an increase in the inference leakage energy in the LL mode compared with the HT mode.

\subsection{System-Level Optimization}

\begin{figure}[t]
    \centering
    \begin{minipage}{0.46\linewidth}
        \centering
        \includegraphics[width=1.0\linewidth]{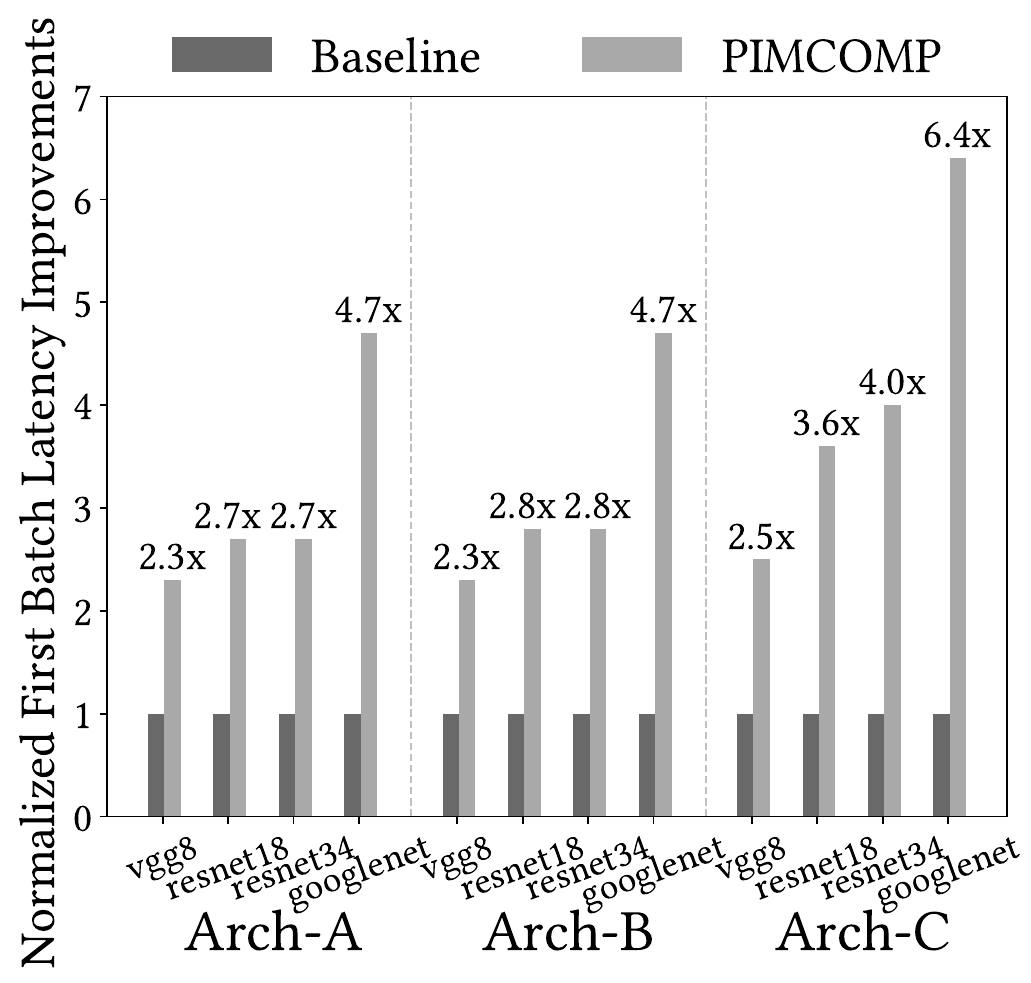}
        \caption{First-batch latency of HT scenario.}
        \label{Fig::HT-latency}
    \end{minipage}
    \centering
    \begin{minipage}{0.46\linewidth}
        \centering
        \includegraphics[width=1.0\linewidth]{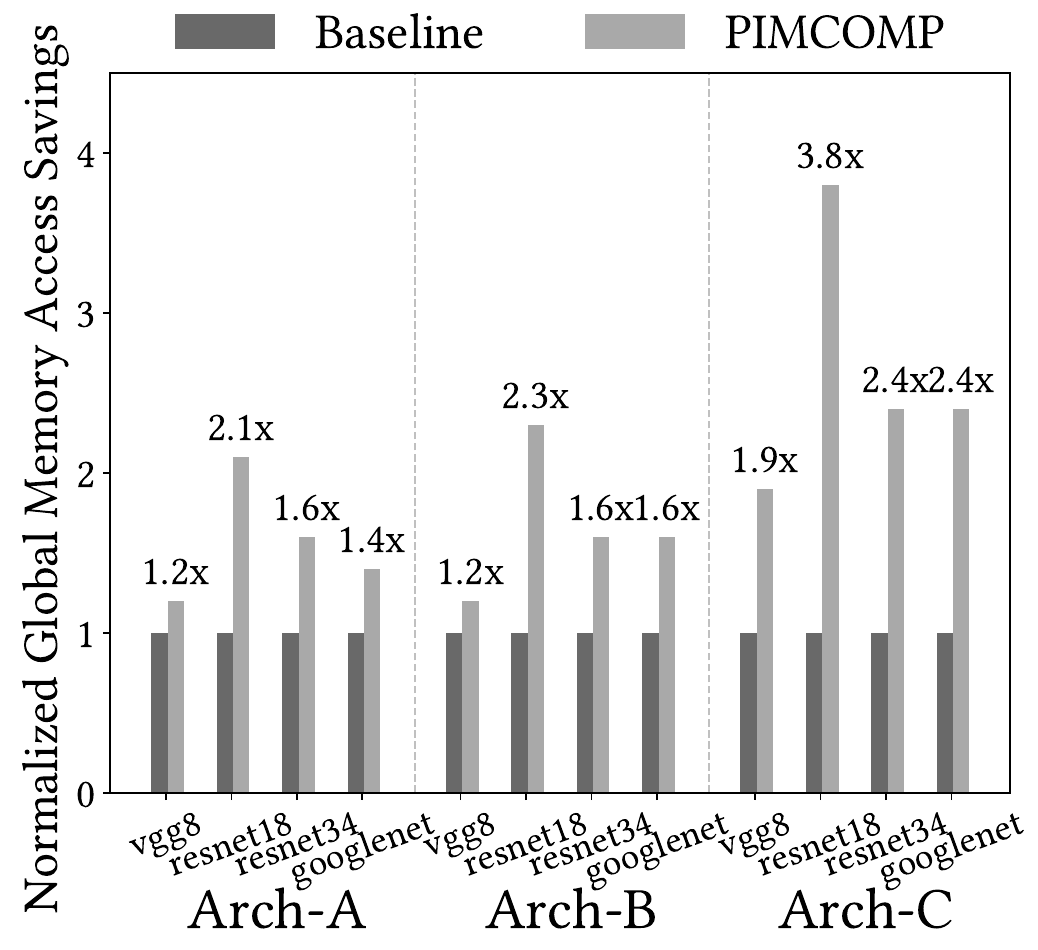}
        \caption{Global memory access volume of HT scenario.}
        \label{Fig::HT-load}
    \end{minipage}
\end{figure}

\begin{figure}[t]
    \centering
    \begin{minipage}{0.46\linewidth}
        \centering
        \includegraphics[width=1.0\linewidth]{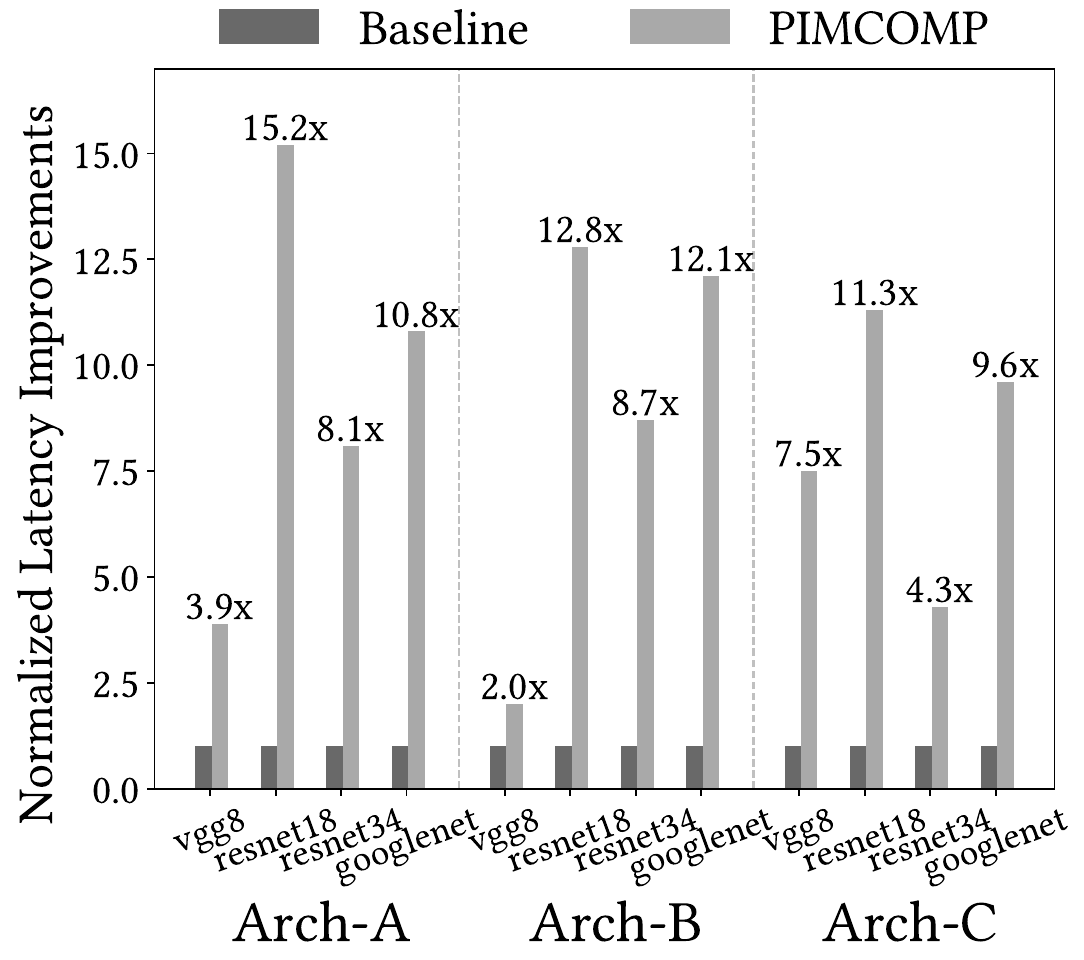}
        \caption{Inference latency of LL scenario.}
        \label{Fig::LL-communication}
    \end{minipage}
    \centering
    \begin{minipage}{0.46\linewidth}
        \centering
        \includegraphics[width=1.0\linewidth]{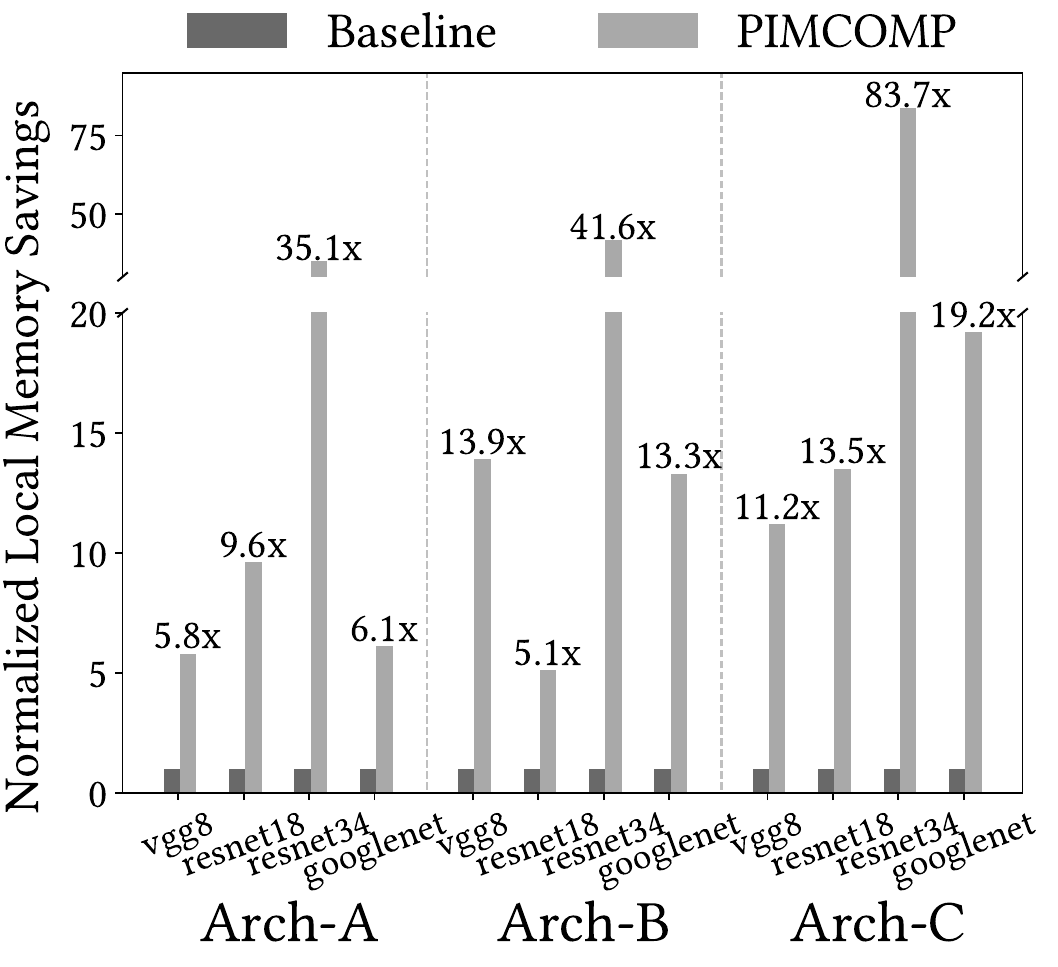}
        \caption{Local memory overhead of LL scenario.}
        \label{Fig::LL-memory}
    \end{minipage}
\end{figure}

\textcolor{black}{Figs.~\ref{Fig::HT-latency} to \ref{Fig::LL-memory} showcase the effects of various individual system-level optimization techniques. Specifically, 
Fig.~\ref{Fig::HT-latency} illustrates the effect of the layer grouping strategy on first-batch latency in HT mode.
Fig.~\ref{Fig::HT-load} shows the impact of the flexible unfolding format, which leverages data reuse features, on global memory access in HT mode.
Fig.~\ref{Fig::LL-communication} highlights the improvement in latency using the heuristic centralized communication mechanism of pixel-level runtime management strategy in the LL scenario. Fig.~\ref{Fig::LL-memory} presents the optimization of local memory requirements within each core using the pixel-level runtime management strategy in the LL scenario.}
The baselines in Figs.~\ref{Fig::HT-latency} to \ref{Fig::LL-memory} represent the performance metrics when \emph{PIMCOMP} does not apply the respective optimization methods. In summary, \emph{PIMCOMP} achieves end-to-end system-level optimization of DNN deployment tasks across dimensions of performance, resource, and memory.

\subsection{Compilation Time}

Table~\ref{Tab::Time} shows the compilation time required by \emph{PIMCOMP} on the Arch-A architecture. The population size in the genetic algorithm is 200, with up to 1000 iterations. The layout-computation mapping occupies most of the time, and the compilation process takes 19.2 minutes on average.
\textcolor{black}{Since compilation is a one-time effort and each model only requires compilation once, the overhead is acceptable.}

\begin{table}[t]
\caption{Compilation time (second).}
\label{Tab::Time}
\centering
\setlength{\tabcolsep}{3.5pt}
\begin{threeparttable}
\begin{tabular}{|c|cc|cc|cc|cc|}
\hline
\multirow{2}{*}{} & \multicolumn{2}{c|}{vgg8}         & \multicolumn{2}{c|}{resnet18}     & \multicolumn{2}{c|}{resnet34}     & \multicolumn{2}{c|}{googlenet}    \\ \cline{2-9} 
                  & \multicolumn{1}{c|}{HT}    & LL   & \multicolumn{1}{c|}{HT}    & LL   & \multicolumn{1}{c|}{HT}    & LL   & \multicolumn{1}{c|}{HT}    & LL   \\ \hline
P\tnote{1}                & \multicolumn{1}{c|}{0.0}   & 0.1  & \multicolumn{1}{c|}{0.1}   & 0.1  & \multicolumn{1}{c|}{0.6}   & 0.6  & \multicolumn{1}{c|}{0.6}   & 0.6  \\ \hline
M\tnote{2}                & \multicolumn{1}{c|}{1034.4} & 414.8 & \multicolumn{1}{c|}{1973.2} & 672.9 & \multicolumn{1}{c|}{1542.1} & 725.5 & \multicolumn{1}{c|}{2170.4} & 633.7 \\ \hline
D\tnote{3}                 & \multicolumn{1}{c|}{3.0}   & 0.5  & \multicolumn{1}{c|}{2.7}   & 0.4  & \multicolumn{1}{c|}{9.5}   & 16.4 & \multicolumn{1}{c|}{7.6}   & 14.7 \\ \hline
Total             & \multicolumn{1}{c|}{1037.4} & 415.4 & \multicolumn{1}{c|}{1976.0} & 673.4 & \multicolumn{1}{c|}{1552.2} & 742.5 & \multicolumn{1}{c|}{2178.6} & 649.0 \\ \hline
\end{tabular}
    \begin{tablenotes}
    \item[] \hspace{-5pt} {\scriptsize $^1$P: layer partitioning. $^2$M: layout-computation mapping.  $^3$D: dataflow scheduling}
    \end{tablenotes}
\end{threeparttable}
\end{table}

\section{Conclusion}
\label{Sec::Conclusion}

We introduce \emph{PIMCOMP}, an end-to-end DNN compiler designed for PIM accelerators. \emph{PIMCOMP} demonstrates excellent versatility, adapting to hardware systems of various scales by abstracting the hardware architecture and the software-hardware interface. \emph{PIMCOMP} integrates a multi-level optimizer for system-level performance and resource optimization to complete deployment efficiently. \emph{PIMCOMP} bridges diverse DNN models and different PIM architectures, promoting the development of the PIM ecosystem.

\bibliographystyle{ieeetr}
\bibliography{PIMCOMP-Ref}

\begin{IEEEbiography}[{\includegraphics[width=1.0in,height=1.1in,clip,keepaspectratio]{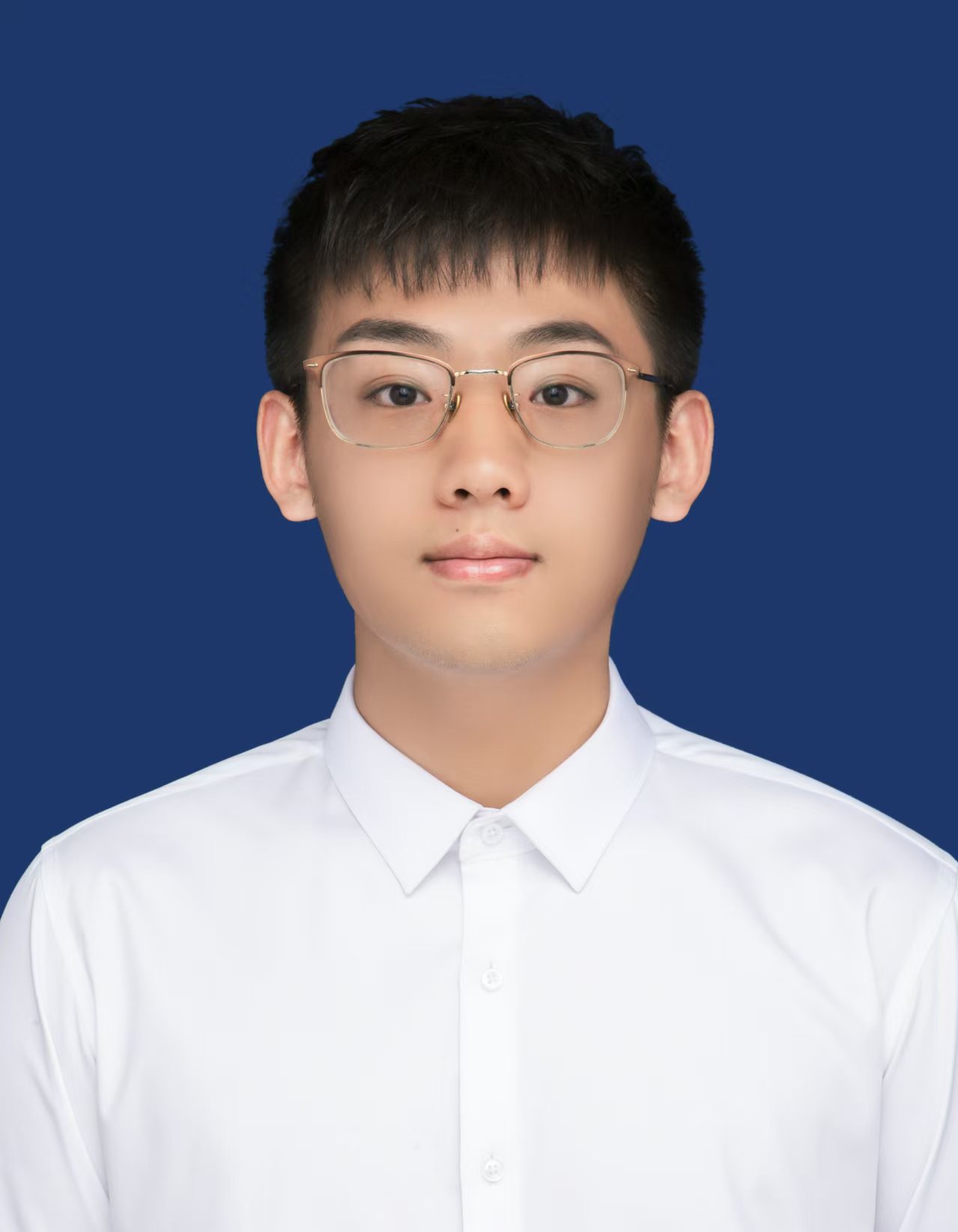}}]{Xiaotian Sun} received the B.S. degree in electronic engineering from Tsinghua University, Beijing, China. in 2021. He is currently pursuing the Ph.D. degree with the Institute of Computing Technology, Chinese Academy of Sciences, Beijing. He is currently interested in accelerating deep neural networks on emerging storage devices.
\end{IEEEbiography}

\begin{IEEEbiography}[{\includegraphics[width=1.0in,height=1.1in,clip,keepaspectratio]{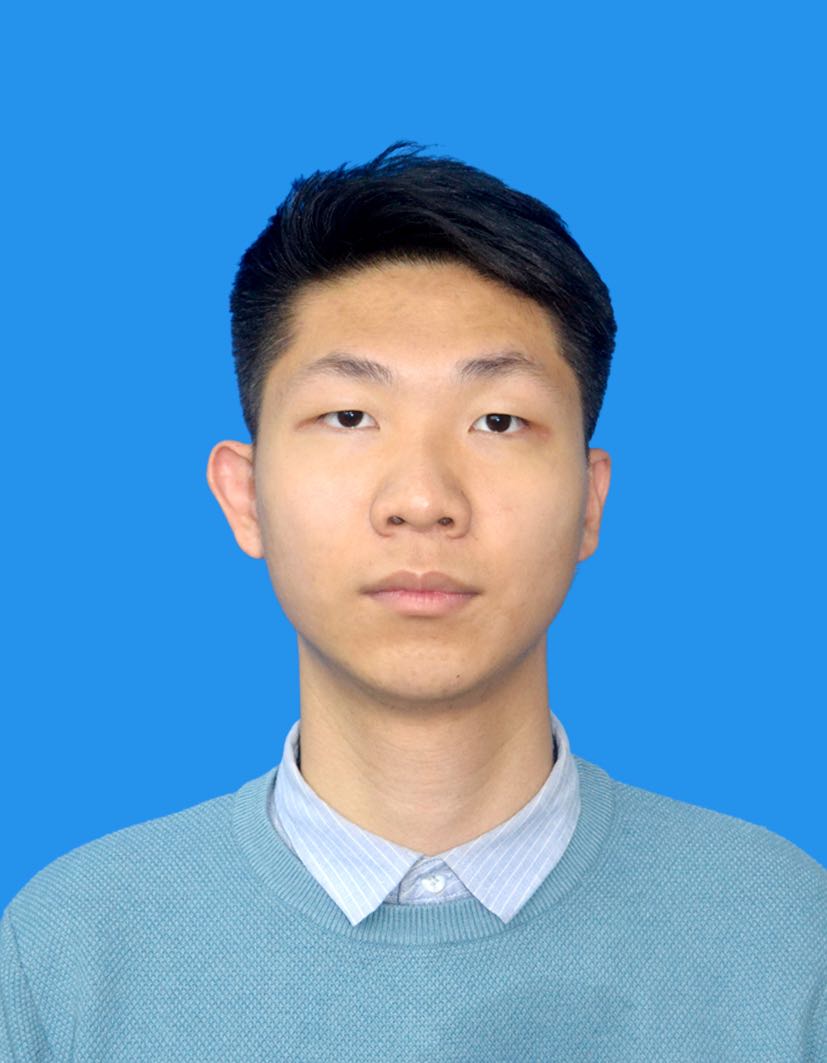}}]{Xinyu Wang} received the B.E. degree from the School of Computer Science and Technology, Shandong University, Jinan, China, in 2022. He is currently pursuing the M.S. degree with the University of Chinese Academy of Sciences, Beijing, China.  His research interests include computer architecture, processing-in-memory, and deep learning.
\end{IEEEbiography}

\begin{IEEEbiography}[{\includegraphics[width=1.0in,height=1.1in,clip,keepaspectratio]{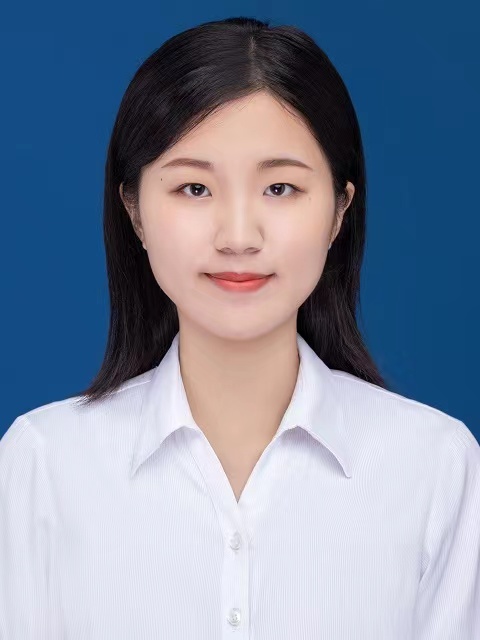}}]{Wanqian Li} received the B.S. degree in Microelectronics Science and Engineering from Nankai University, Tianjin, China. in 2020. She is currently pursuing the Ph.D. degree with the Institute of Computing Technology, Chinese Academy of Sciences, Beijing. She is currently interested in automatic synthesis of processing-in-memory architectures and acceleration of large language model inference.
\end{IEEEbiography}

\begin{IEEEbiography}[{\includegraphics[width=1.0in,height=1.1in,clip,keepaspectratio]{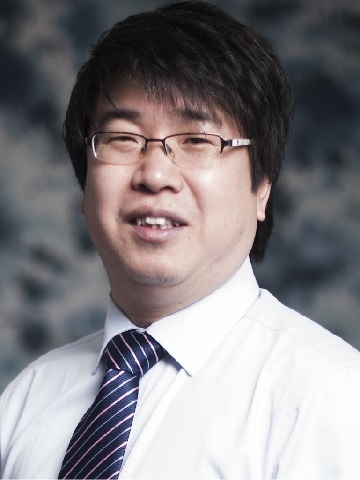}}]{Yinhe Han} (Member, IEEE) received the M.S. and Ph.D. degrees in computer science from the Institute of Computing Technology (ICT), Chinese Academy of Sciences (CAS), in 2003 and 2006, respectively. He is currently a Professor with ICT, CAS. His main research interests are microprocessor design, integrated circuit design, and computer architecture.
\end{IEEEbiography}

\begin{IEEEbiography}[{\includegraphics[width=1.0in,height=1.1in,clip,keepaspectratio]{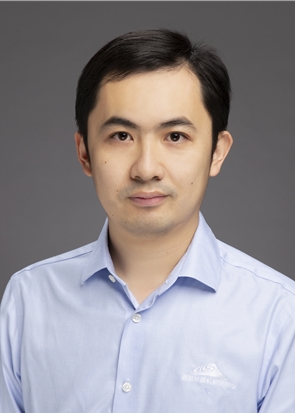}}]{Xiaoming Chen} (Member, IEEE) received the B.S. and Ph.D. degrees in electronic engineering from Tsinghua University, Beijing, China, in 2009 and 2014, respectively. He is currently a Professor with the Institute of Computing Technology, Chinese Academy of Sciences, Beijing. His current research interest is focused on design automation for integrated circuits and  PIM architectures.
\end{IEEEbiography}

\end{document}